\documentclass[fleqn,usenatbib]{mnras}

\usepackage{newtxtext,newtxmath}

\usepackage[T1]{fontenc}

\DeclareRobustCommand{\VAN}[3]{#2}
\let\VANthebibliography\thebibliography
\def\thebibliography{\DeclareRobustCommand{\VAN}[3]{##3}\VANthebibliography}

\usepackage{graphicx}
\usepackage{float}	
\usepackage{amsmath}	
\usepackage{multirow,bigdelim}

\usepackage{colortbl}
\defcitealias{Enzi25}{E25}

\defcitealias{2024MNRAS.528.7564B}{B24}

\usepackage{xcolor}

\usepackage{tikz}
\usetikzlibrary{shapes.geometric, arrows, backgrounds, positioning, fit}

\title[Gaussian processes on RTU grids]{Gaussian processes on ray-guided transformed uniform grids for fast, flexible, and auto-differentiable adaptive source reconstruction in lens modelling}

\author[W. J. R. Enzi et al.]{
Wolfgang J. R. Enzi,\thanks{E-mail: wolfgang.enzi@port.ac.uk}
Coleman M. Krawczyk, Tian Li, 
and Thomas E. Collett
\\
Institute of Cosmology and Gravitation, University of Portsmouth, Burnaby Rd, Portsmouth PO1 3FX, UK
}

\date{}

\pubyear{2026}

\begin{document}
\label{firstpage}
\pagerange{\pageref{firstpage}--\pageref{lastpage}}
\maketitle

\begin{abstract}
Strong gravitational lensing constrains cosmology and dark matter, but robust inference requires accurate source reconstruction. The achievable source resolution is highly position-dependent. Adaptive meshes can place resolution where needed, but typically rely on discontinuous operations, such as Delaunay tessellations or Voronoi binning, which can restrict regularization choices and break differentiability.
%
In this paper, we present a novel approach for modelling the source on a ray-guided transformed uniform grid (RTU grid), that is adaptive to the lens mass model, auto-differentiable and  flexible with respect to the regularization by allowing for an arbitrary choice of power spectrum. 
%
We achieve this by defining the source as a Gaussian process on a uniform grid, which is then transformed based on the cumulative distributions of rays traced back to the source plane. This approach ensures that source pixels contain a more uniform number of rays. The approach is fast by leveraging the fast Fourier transform to describe the Gaussian process in Fourier space.
%
We apply this new approach to mock data and show that it achieves comparable fit quality with fewer source pixels, typically corresponding to about a factor of two fewer pixels per dimension, and increases Evidence Lower Bounds (ELBOs) for the same number of pixels. Using the RTU grid only mildly affects the difference in ELBO for models with and without substructures within lens galaxies.
%
A fast, flexible, and auto-differentiable source reconstruction can greatly benefit the analysis of large samples of lens systems, e.g. those found within the Euclid survey.
\end{abstract}




\section{Introduction}

Strong gravitational lensing (SGL) is a powerful probe of cosmology and astrophysics. It is used to constrain the Hubble parameter $H_0$ \citep[e.g.][]{2024SSRv..220...48B}, to test general relativity \citep[e.g.][]{2018Sci...360.1342C}, and to constrain the distribution of dark matter on galactic and sub-galactic scales \citep[e.g.][]{2024SSRv..220...58V}. In strong gravitational lensing, the light of a background source (e.g. a galaxy) is deflected by a mass in the foreground (e.g. another galaxy) to such a degree that multiple distorted images of the source can be observed. In all these analyses, the source has to be simultaneously reconstructed with the light and mass distribution of the lens. Robust constraints on cosmology or the mass distribution will therefore depend on the priors that determine the reconstructed source \citep{2022MNRAS.510.2480D,2024arXiv240712910D,2024MNRAS.528.7564B}. With the increasing availability of high-quality high-resolution data (e.g. from the James Webb Space Telescope), source model assumptions will have a significant impact on the reconstructed lens parameters and the biases affecting cosmology and dark matter constraints \citep[see e.g.][]{2024A&A...692A..87G}.

Priors on the reconstructed source balance freedom with realism. Too little freedom in the source leads to poor fidelity in fitting the data or overestimation of the abundance of structure in the lens mass distribution. Too much freedom, on the other hand, leads to sources that do not resemble observed galaxies and are physically implausible. Recently,  \citet[][]{2025arXiv251119595L} trained generative models on the shapes of nearby galaxies to improve realism of the inferred source galaxies by generating their features according to what is observed in nearby galaxies with high-resolution imaging. 

Free-form source models, on the other hand, infer the source brightness on regular or irregular grids in the source plane \citep[see e.g.][]{2003ApJ...590..673W}. Common choices for regularization are curvature and gradient priors, which aim to suppress excessive structure within sources when not required by the data \citep[][]{2006MNRAS.371..983S}. Other regularization approaches, based on Shapelet or Wavelet decompositions, and Gaussian processes have also been proposed to regularize reconstructions, as they allow for a more flexible structure in the reconstructed source during inference \citep[][]{2022MNRAS.512..661K,2022A&A...668A.155G,2024A&A...682A.146R}. In the case of Gaussian processes, the limitation is that fast Fourier transforms (FFTs) require regular grids. 
This creates a practical tension: Gaussian-process priors provide a flexible way to regularize source reconstructions, but their efficient implementation relies on regular grids, whereas the information content of strong-lensing data is highly non-uniform across the source plane. 

The resolution at which the source can be reconstructed varies over the source plane for two reasons: First, the parts of the source that create four images will show approximately twice the sampling rate of those that create only two images. Second, the magnification across the image plane corresponds to a continuous change of the sampling rate over the source plane. To account for this variation, many lens modelling codes often employ adaptive meshes \citep[see e.g.][]{2009MNRAS.392..945V,2010MNRAS.408.1969V,2015MNRAS.452.2940N}. These adaptive meshes are typically Delaunay (or Voronoi) tessellations. The source brightness is interpolated between the values at the vertices (or in the cells) of these meshes.

 This can often lead to discontinuities in the likelihood function, since parts of the mesh can change in a discrete fashion when mass models change, e.g. the edges in a Delaunay mesh may swap the vertices they connect \citep[see e.g. Figure B1 of][]{2024MNRAS.528.7564B}. Nearest-neighbour interpolation methods \citep[see e.g.][]{2024MNRAS.52710480N} can alleviate some of these discontinuities but are computationally more costly.

In this paper, we present a novel  approach to source reconstruction that combines these two requirements: the computational structure needed for fast Gaussian-process regularization via FFTs and the adaptive resolution required by the non-uniform lensing information. We achieve this by modelling sources on a uniform grid, which is transformed based on the cumulative distributions of rays traced back to the source plane. This  ray-guided transformed uniform grid (RTU grid) effectively provides an adaptive grid that is continuously differentiable with a flexible and fast regularization. We present this approach in the forward modelling framework of {\sc Herculens} \citep[][]{2022A&A...668A.155G}. Herculens is written using the auto-differentiable programming package  {\sc Jax} \citep[][]{jax2018github} and as such is able to take advantage of state-of-the-art sampling algorithms implemented in {\sc Numpyro} \citep[][]{phan2019composable,bingham2019pyro}. The approach can be easily implemented for inverse or hybrid lens modelling frameworks, including \citep[e.g. pyautolens,][]{2021JOSS....6.2825N}, where the source is modelled through the inversion of large matrices. 

The remainder of this paper is structured as follows:  Section \ref{sec:source_models} describes the novel source reconstruction approach. Section \ref{sec:mock_generation} describes the mock data and real data that we use to test this approach. Section \ref{sec:Analysis} outlines the inference approach we use throughout this work. Section \ref{sec:Results} presents the results of our analyses and compares them to the uniform grid approach previously implemented in {\sc Herculens}. Section \ref{sec:Discussion} discusses possible extensions and the limitations of this approach. Finally, Section \ref{sec:Conclusion} provides an outlook on how this source reconstruction can benefit the modelling of large samples of lens systems in the near future.

\section{Modelling the source on Ray-guided Transformed Uniform grids}
\label{sec:source_models}
In the following Section, we describe source models on a uniform grid and the novel RTU grid.
A short description of the lens light and mass models is provided in Appendix \ref{sec:StrongGravitationalLensing}.
\subsection{The old approach: Source reconstruction on uniform grids} 

 Our uniform-grid free-form source brightness follows the approach outlined in \citet[][]{Enzi25}, which models the source as a Gaussian process. The source brightness value in each pixel of the uniform grid is the component of a vector sampled from a multivariate Gaussian distribution. The extent of this grid is determined by the size of the smallest square box containing all the rays that were included in a mask around the arcs. The covariance structure is assumed to be diagonal in Fourier space and determined by a power spectrum $\vec P$. A random realization of this field, $\vec f$, can be generated by drawing a field of white noise from a zero-mean unit-variance Gaussian distribution and then pointwise multiplying ($\odot$) its Fourier transform, $\vec \xi$, with the square root of the power spectrum: 
 \begin{equation} \vec f= \mathcal{F}^{-1} \left(\sqrt{\vec P} \odot \vec \xi \right) \,, \end{equation}
 where $\mathcal{F}^{-1}$ is the inverse Fourier transform. Since the grid on which we define our source is regular, we can apply the FFT that is already implemented in {\sc JAX}. 
 Positivity of the source is enforced through a softplus function $\phi_{\rm sp}$. The final source brightness values are: 
 \begin{equation} \phi_{\rm sp}(\vec x ) = \log(1+\exp(\vec x))  \end{equation}
 and 
 \begin{equation} \vec s_{\rm source} = \left( \phi_{\rm sp}(\vec f / s_0) \times s_0 \right)^{\alpha_{\rm src}}\,, \end{equation}
 with $s_0=0.01$ being the scale over which the softplus function transitions from asymptotically exponential to linear growth and the power $\alpha_{\rm src}$ enabling sharper features in the source. The surface brightness outside the pixel grid is set to zero. We use a bi-linear interpolation to evaluate the source brightness in between pixels, defining $s_{\rm source}(\vec x )$ for all positions $\vec x$.

 Several choices can be made for the power spectrum that effectively describes the regularization prior. A particularly flexible choice is the Mat\'ern power spectrum \citep[see e.g.][]{2020MNRAS.499.5641V,2022MNRAS.516.1347V,2024A&A...682A.146R,Enzi25}: 
\begin{equation} P_s(k) = \sigma_{\rm src}^2 \times 4 \pi n_{\rm src} \times\left(\frac{2n_{\rm src}}{\zeta_{\rm src}^2}\right)^{n_{\rm src}}\times \left(\frac{2n_{\rm src}}{\zeta_{\rm src}^2} + k^2 \right)^{-(n_{\rm src}+1)}\,, \end{equation} 
which reduces to the curvature (gradient) regularization if $n_{\rm src} = 1$ ($n_{\rm src}=0$) for large $\zeta_{\rm src}$. The amplitude of fluctuations $\sigma^2_{\rm src}$, the power index $n_{\rm src}$, and the characteristic scale $\zeta_{\rm src}$ are left to vary freely. We infer $\alpha_{\rm src}$ together with the power spectrum parameters.

\begin{figure*}
    \centering
    \includegraphics[width=0.49\textwidth]{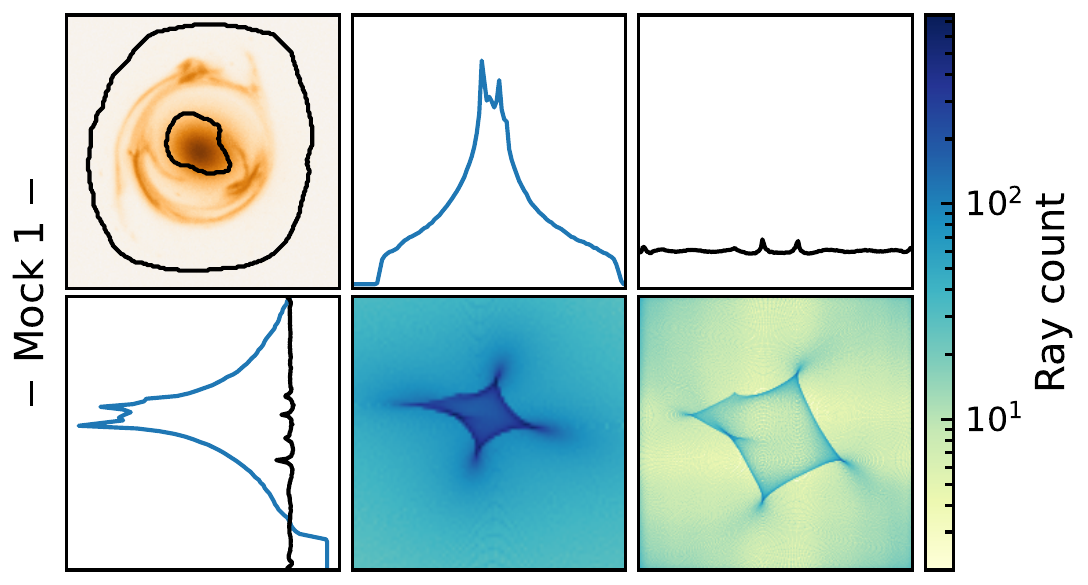} 
    \includegraphics[width=0.49\textwidth]
    {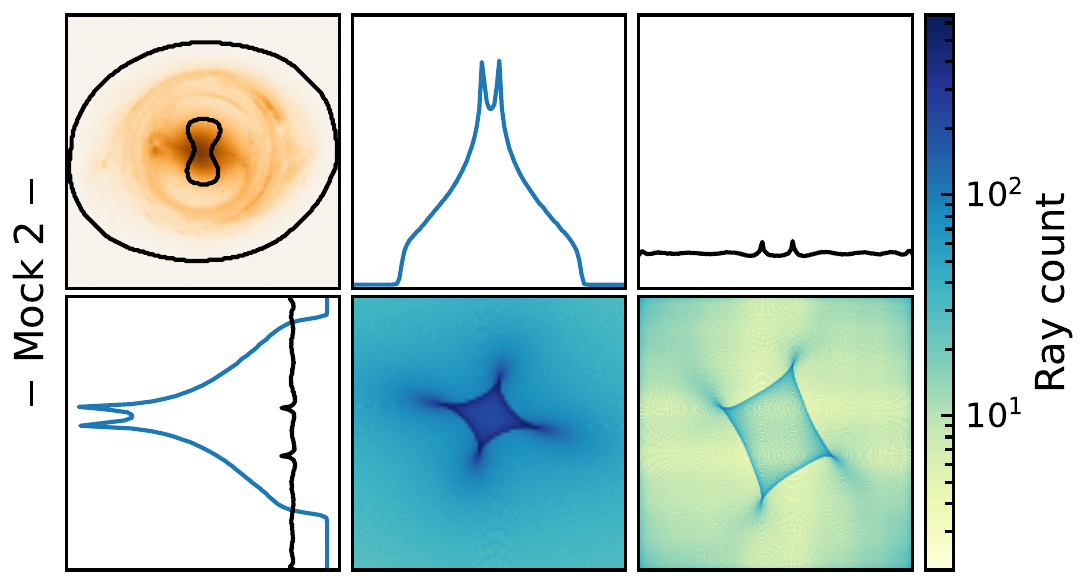}\\
    \includegraphics[width=0.49\textwidth]{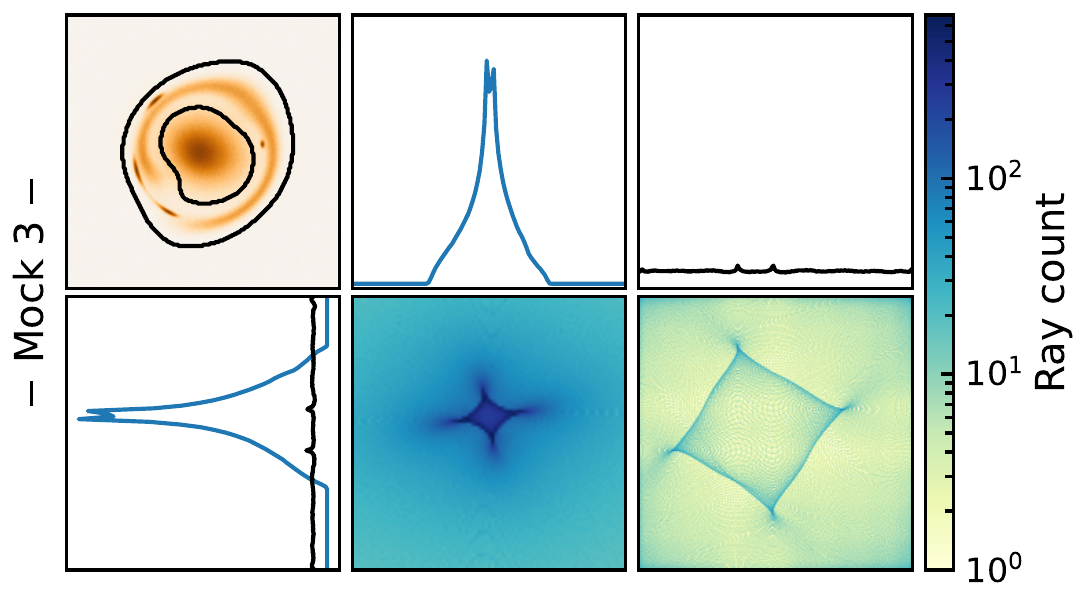}
    \includegraphics[width=0.49\textwidth]{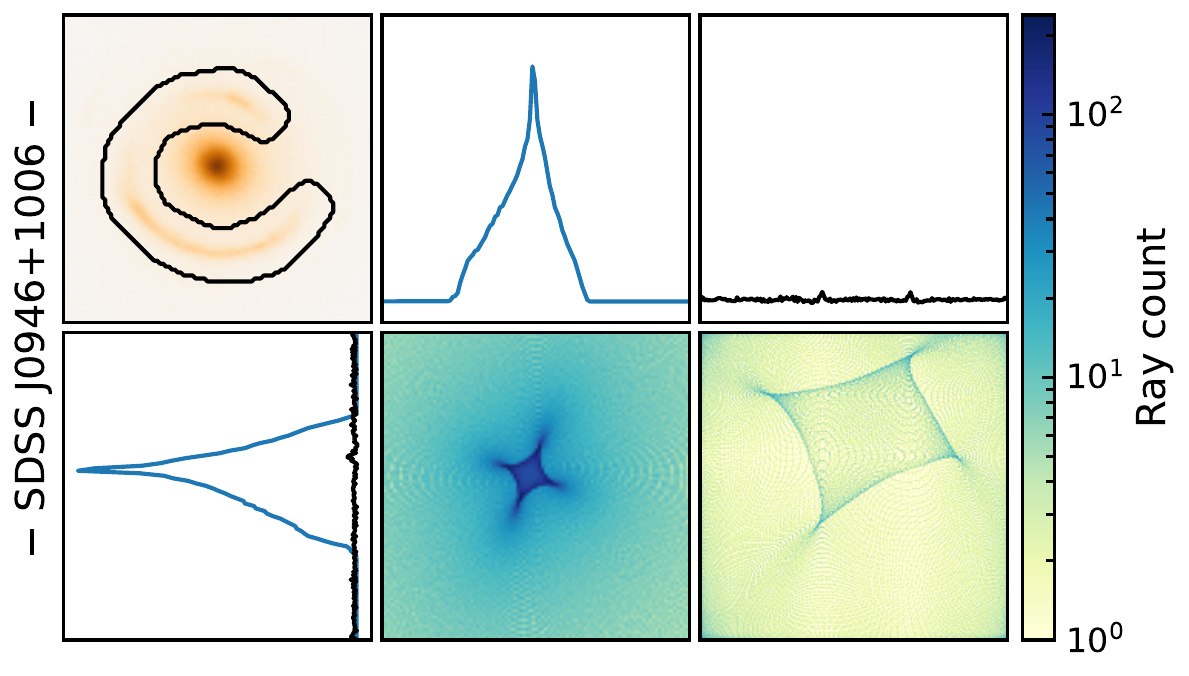}
    \caption{  
    This figure shows, for four test cases, the non-uniform ray density created by lensing and how the RTU grid provides a more even sampling. The four test cases are shown in the four subfigures for Mock 1 (top left),  Mock 2 (top right),  Mock 3 (bottom left), and SDSS J0946+1006 (bottom right). Each subfigure contains the following panels: the data with the arc mask shown as a black contour (top left), the marginal distribution of rays along one dimension for the uniform grid (top middle) and the RTU grid (top right), the marginal distribution of rays along the other dimension for the uniform and RTU grids (bottom left), the 2D ray-density histogram in the source plane (bottom middle), and the 2D ray-density histogram on the RTU grid (bottom right). To highlight that the source plane is sampled more uniformly by the rays within the arc mask after the transform, we show the RTU-grid marginal histograms only for those rays. In the case of SDSS J0946+1006, we show the ray density of the highest-$\log(\mathrm{ELBO})$ model inferred from the data, as described in Section \ref{sec:Results}.
    }
   
    \label{fig:Mocks}
\end{figure*}

\subsection{The new approach: Source reconstruction on RTU grids}

The previously described uniform grid is inefficient in lens modelling, because the ray density in the source plane is not uniform due to changes in magnification and image multiplicity (see Figure \ref{fig:Mocks}). Source pixels that contain many rays may indicate a pixel size that is too large to resolve source features. A smaller pixel size and more pixels can alleviate this, but comes with increasing computational cost. To alleviate this problem, we propose a transformation of the coordinates $\vec x$ to new coordinates $\vec y = \vec g(\vec x)$ that creates an approximately uniform ray density in the new coordinates. Having a uniform grid in the new coordinates will redistribute resolution to the places where it is required most, thereby rendering the inference problem more computationally efficient. In the following paragraphs, we will describe how to create this transformation. Due to the uniform grid in the new coordinates, we can still take advantage of the flexibility of Gaussian processes and their computational speed through the use of FFTs. 

We start by considering the empirical 1D cumulative density (eCDFs) of rays in the source plane. We obtain the eCDFs by projecting ray positions onto each coordinate axis separately and then sorting them. We then fit cubic splines to these eCDFs to obtain the desired transformation in each dimension, i.e. $\vec y = \vec g(\vec x) = ( \text{eCDF}_1(x_1),\text{eCDF}_2(x_2) )$.\footnote{We explain the fitting procedure in more detail in the Appendix \ref{sec:fitting}, since we employ several techniques to improve its speed while keeping our approximation stable and accurate.} This transformation is equivalent to constructing a separable copula via marginal eCDFs \citep[see e.g.][]{nelsen2007introduction}.
We choose cubic splines to guarantee that the transformation is at least twice continuously differentiable.

Figure \ref{fig:Mocks} shows histograms of the rays traced to the source plane before and after the transformation for the four test cases described in Section \ref{sec:mock_generation}. To generate a roughly uniform ray density, the area within the diamond caustic is increased by  a factor of approximately 2-4, as expected with the image multiplicity changing by the same amount. This approach is only approximate since we assume separability of the two relevant dimensions.\footnote{This is not exact because of the additional changes through magnification.} This assumption is not exact, but our mock examples show that it captures the dominant variation in the ray density in the case of galaxy-galaxy lensing. Due to the smoothness of our spline fit the constraints near caustics will remain as local peaks in the transformed ray density marginals (see Figure \ref{fig:Mocks}). A more explicit discussion of these features is provided in Appendix \ref{sec:fitting}.

The RTU grid should cover all rays that map to the observed arcs. To avoid allocating source-plane resolution to regions where the source is dark, we construct the eCDFs using only image-plane rays that lie within an arc mask, as shown in Figure \ref{fig:Mocks}. \footnote{The selection of rays is done before any super-sampling is applied.} 
We note that we do not remove any light of the model arcs that falls outside of this mask in the observed plane.
Assuming that $s^{*}_{\rm source}(\vec y)$ was generated as described in the previous section, we transform the positions according to our mapping: \begin{equation} s_{\rm source}(\vec x)= s^{*}_{\rm source}(\vec g(\vec x)) \,, \end{equation} The computational time for source generation is proportional to the number of pixels both in the transformed and regular uniform grid case. The transformation is recalculated every time the mass parameters change according to the resulting density of rays covering the source plane.

\subsection{Rule of thumb for pixel number}

The proposed approach further provides a rule of thumb choice for the maximum number of grid pixels, which is known at the onset of the problem. The density of rays determines how well the source is constrained at each location. If the density of rays is too low, then the prior will dominate the reconstruction.  
Choosing about 4 rays per pixel is a conservative approach because it aims to give each pixel the number of rays that match the locally free parameters of the bi-linear interpolation between brightness values.
Throughout this work, $N_{\rm src}$ denotes the number of source-plane pixels
along one dimension, so that the total number of source pixels is
$N_{\rm src,tot}=N_{\rm src}^2$. Analogously, for image pixels we choose $N_{\rm img,tot}=N_{\rm img}^2$.
If one chooses a super-sampling factor of $s_{\rm super}$, the resulting pixel grid number along each dimension is simply $N_{\rm src} \approx \frac{ N_{\rm img} \cdot s_{\rm super}}{2}$.\footnote{When the arc mask is included, we consider the corresponding total number of pixels within this mask, $N_{\rm mask, tot}^*$. In this case the estimated ideal number  of source pixels becomes $N_{\rm src} \approx \frac{  \sqrt{N_{\rm mask, tot}^*} \cdot s_{\rm super}}{2}$.} Increasing the number of pixels much further will lead to interpolation through the prior correlation structure.

\begin{figure*}
    \centering
    \includegraphics[width=\textwidth]{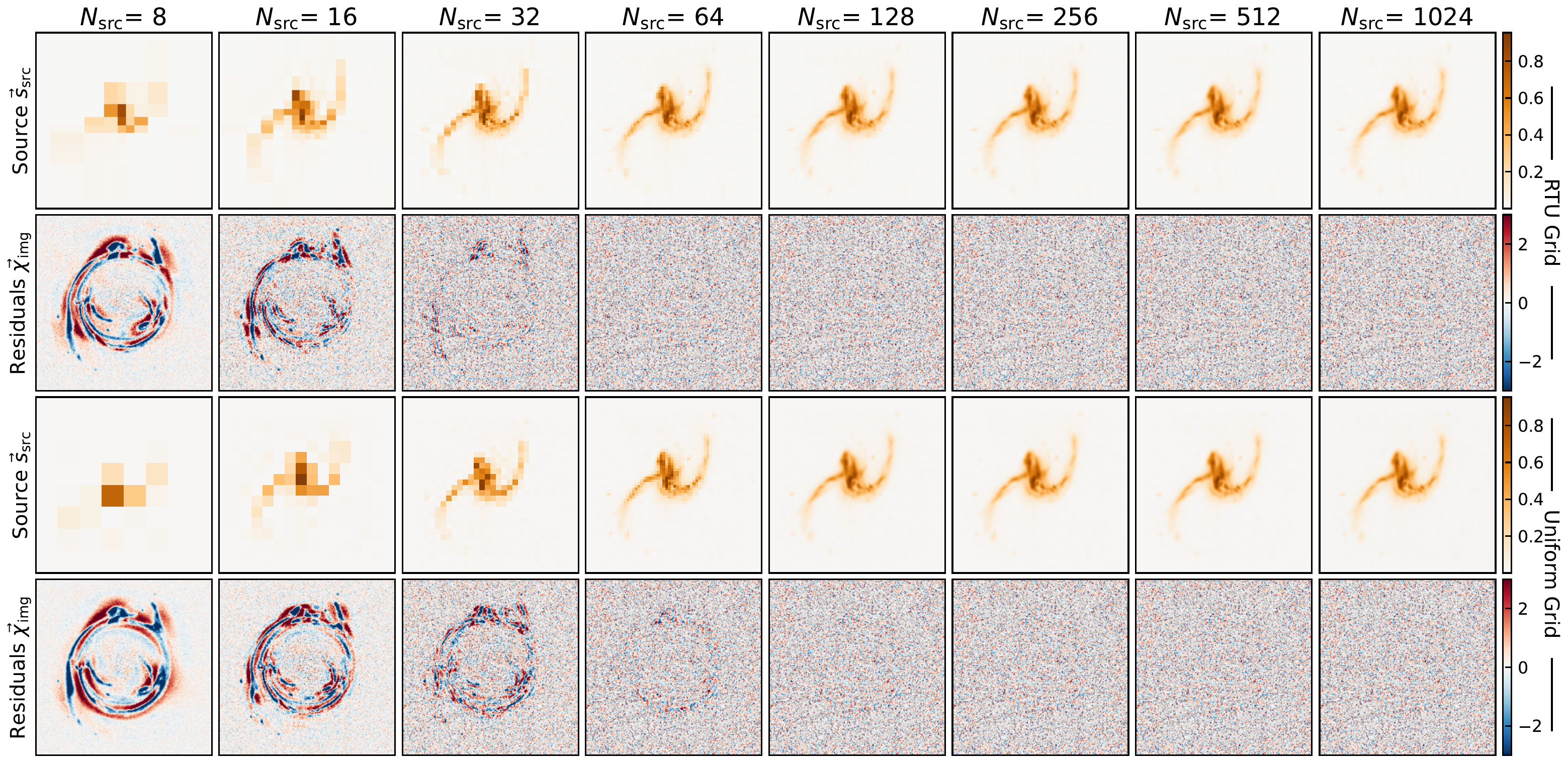} 
    \caption{This figure shows the reconstructed sources and normalized residuals as a function of the pixel number per side, $N_{\rm src}$, for the RTU grid (top two rows) and the uniform grid (bottom two rows).  Residuals are clipped to a range of $\pm3 \sigma$. From left to right, number of pixels per side is $N_{\rm src} \in \{8, 16, 32, 64, 128, 256, 512, 1024\}$. Pixels outside of the source grids are set to 0. All panels show the results of the SVI chains with the highest $\log(\mathrm{ELBO})$.  The shown models contain an NFW substructure.  }
    \label{fig:ReconstructionMock1}
\end{figure*}

\begin{figure*}
    \centering
     \includegraphics[width=\textwidth]{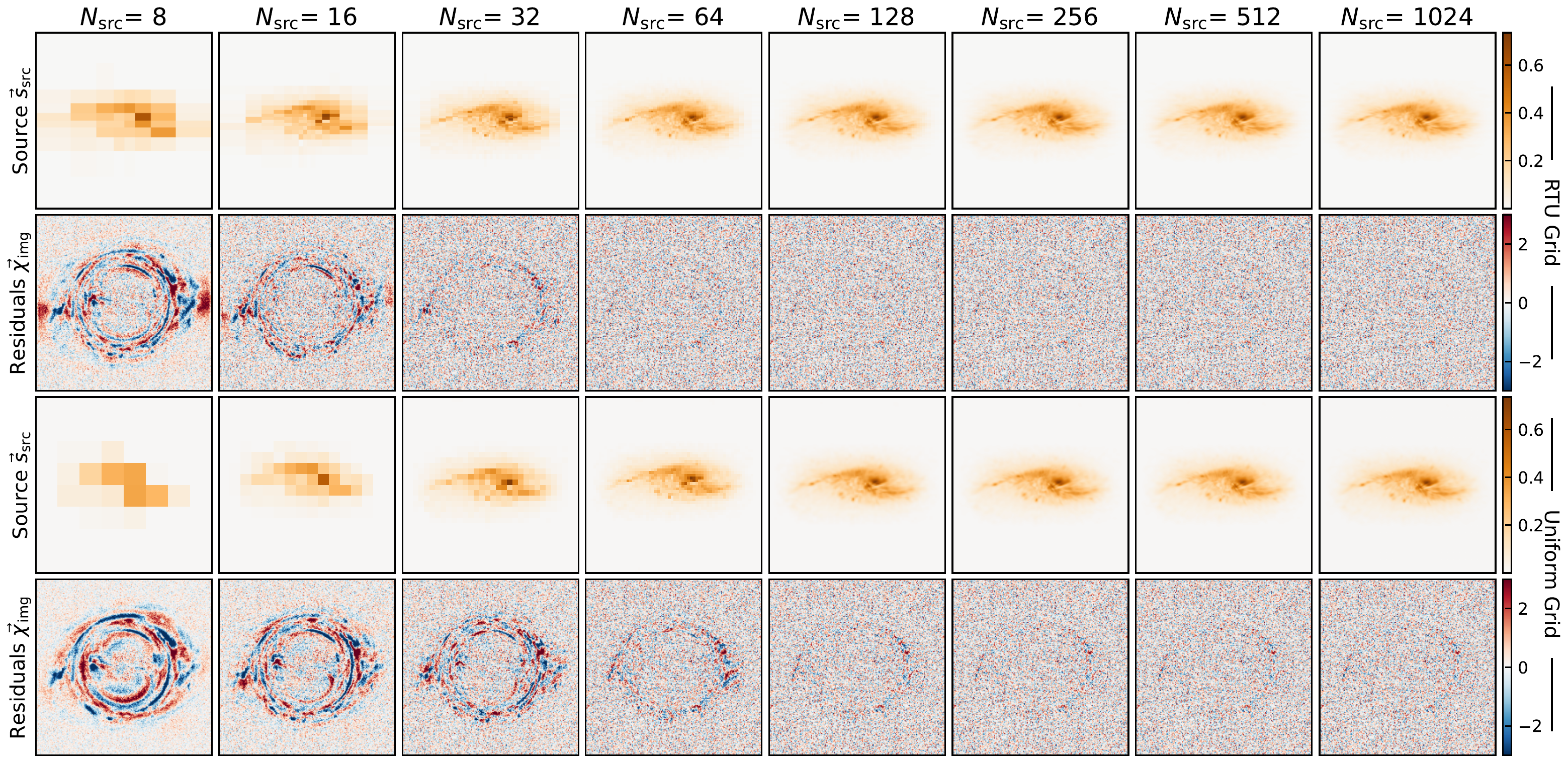} 
    \caption{Same as Figure \ref{fig:ReconstructionMock1} but for Mock 2. This figure shows models without a substructure. }
    \label{fig:ReconstructionMock2}
\end{figure*}

\begin{figure*}
    \centering
    \includegraphics[width=\textwidth]{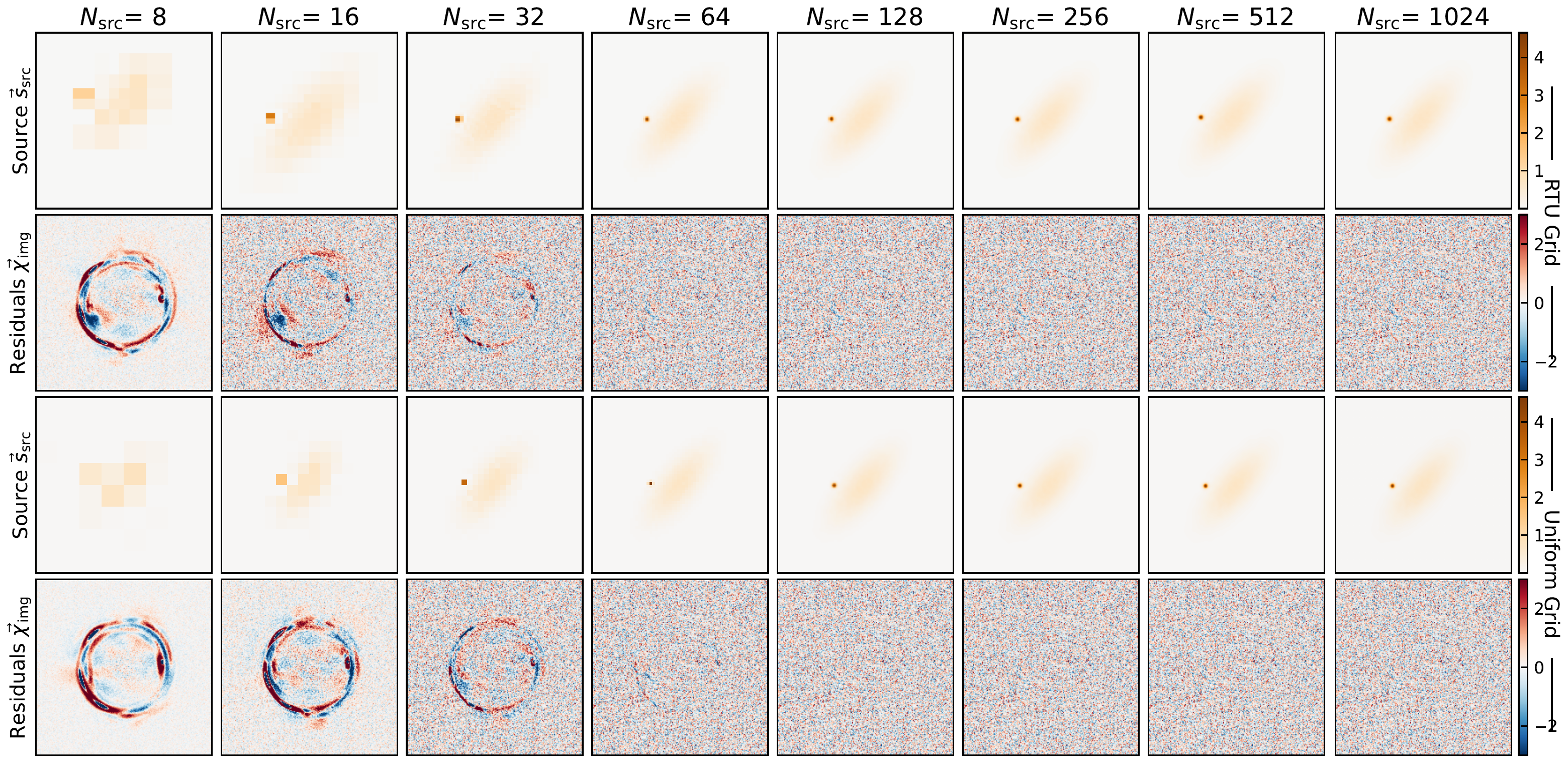} 
    \caption{Same as Figure \ref{fig:ReconstructionMock1} but for Mock 3. This figure shows models containing an NFW substructure.}
    \label{fig:ReconstructionMock3}
\end{figure*}

\begin{figure*}
    \centering
    \includegraphics[width=\textwidth]{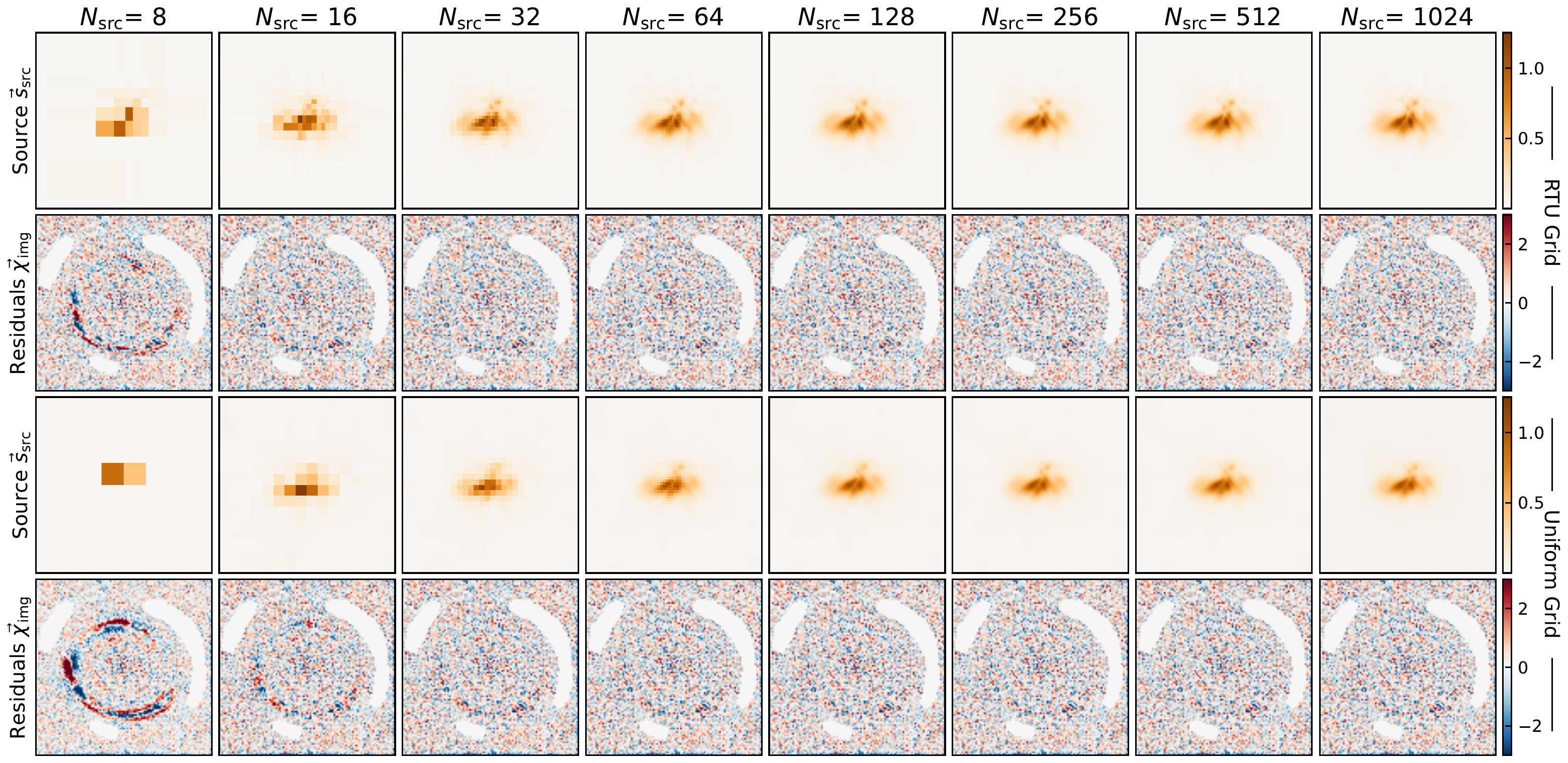} 
    \caption{Same as Figure \ref{fig:ReconstructionMock1} but for SDSS J0946+1006. This figure shows the model containing a substructure. }
    \label{fig:ReconstructionJackpot}
\end{figure*}

\begin{figure}
    \centering
    \includegraphics[width=\hsize]{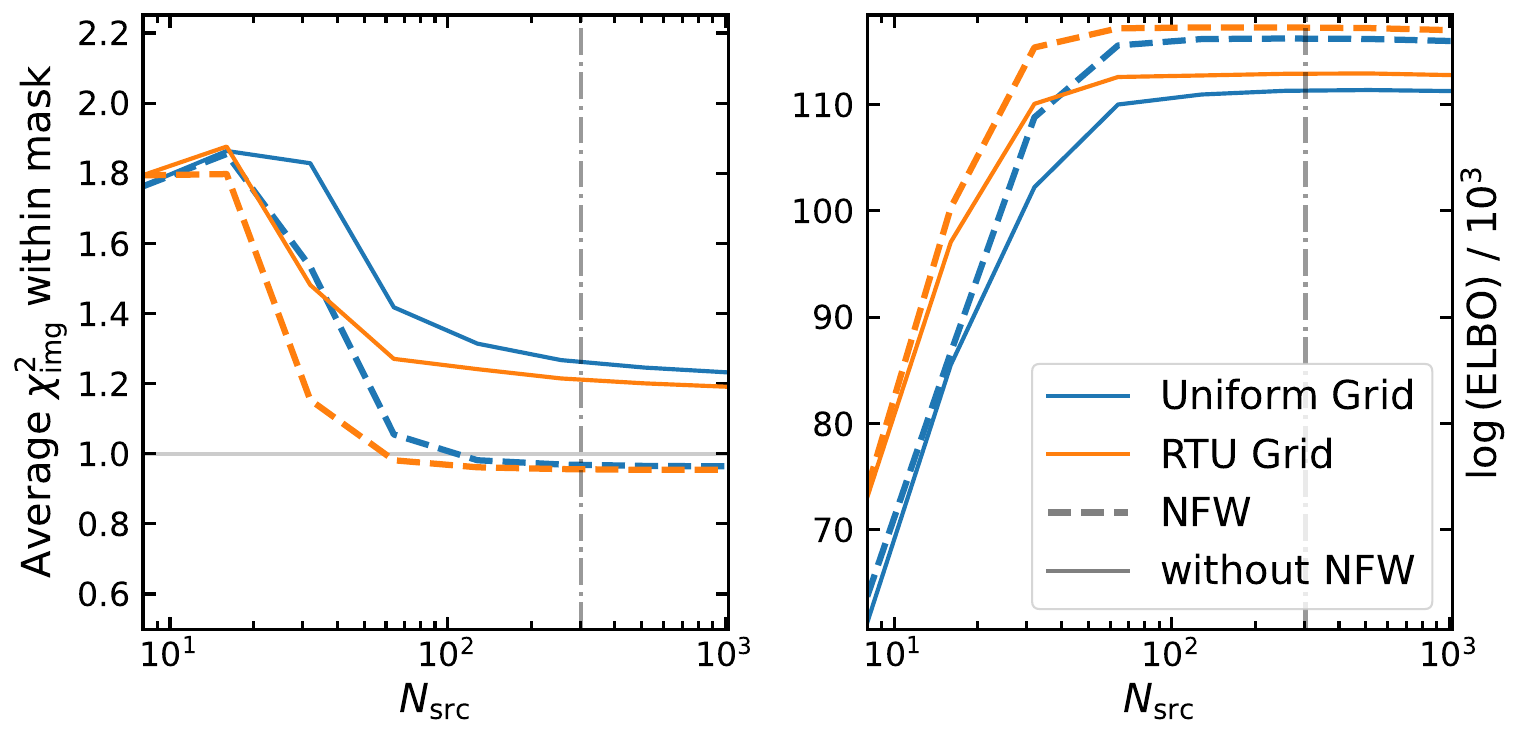} \\
    \includegraphics[width=\hsize]{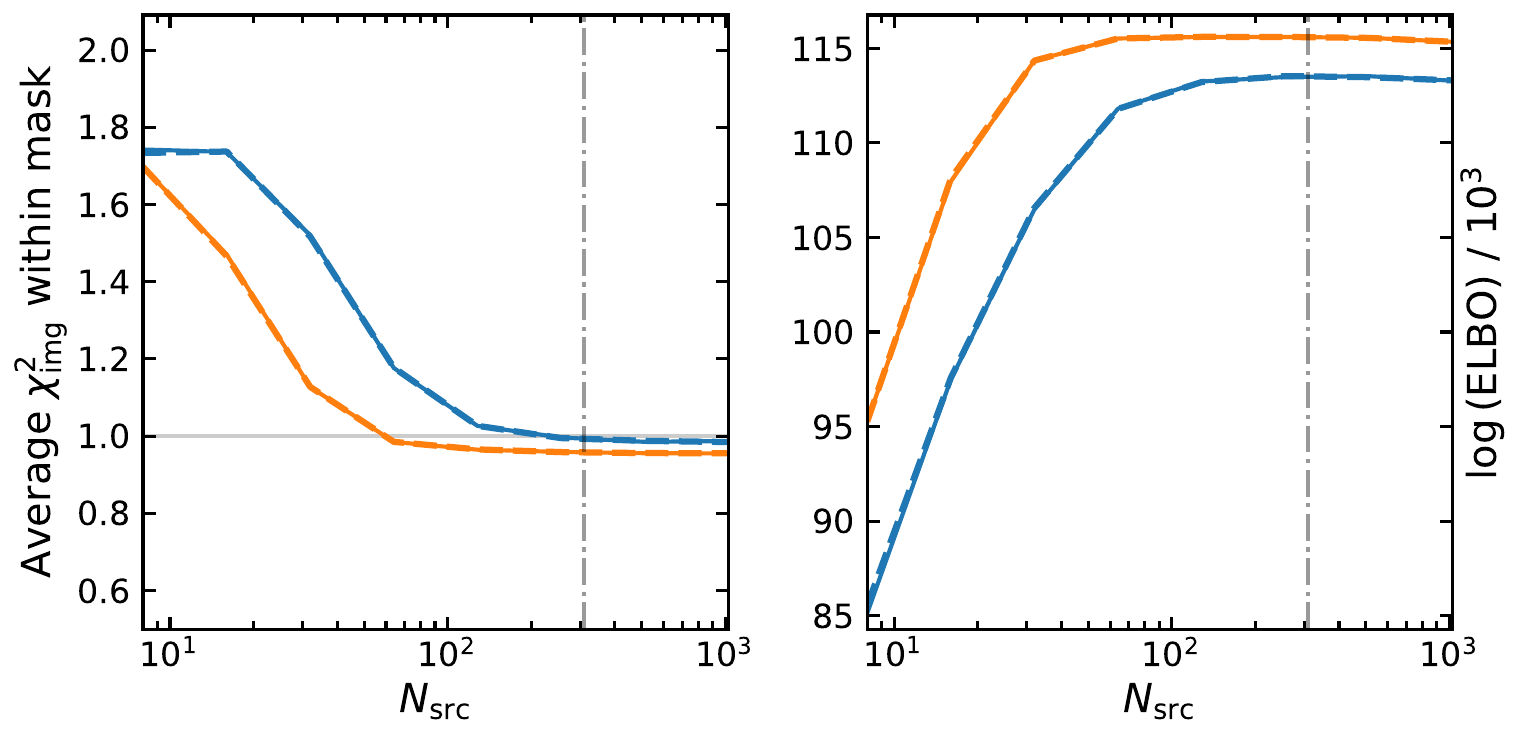} \\
    \includegraphics[width=\hsize]{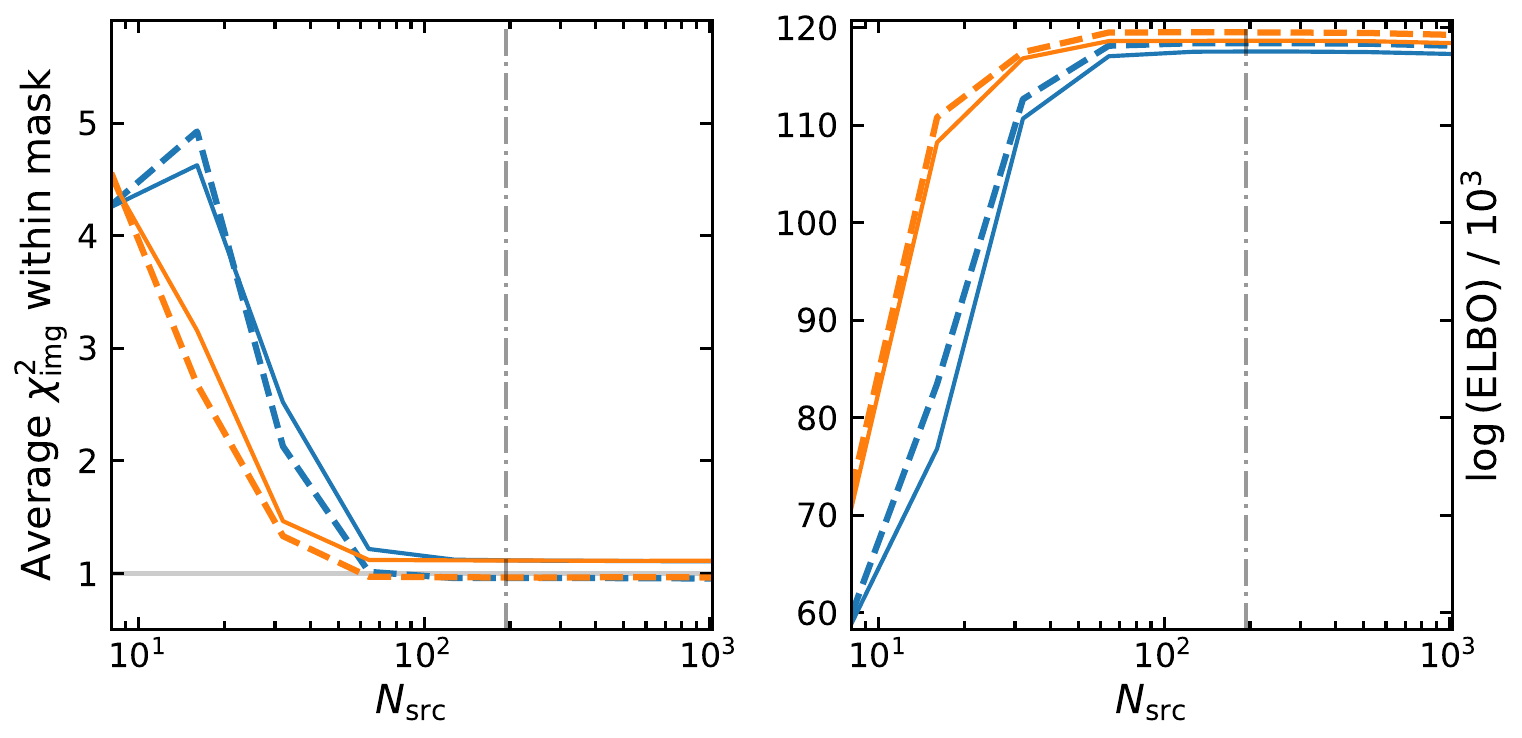} \\
    \includegraphics[width=\hsize] {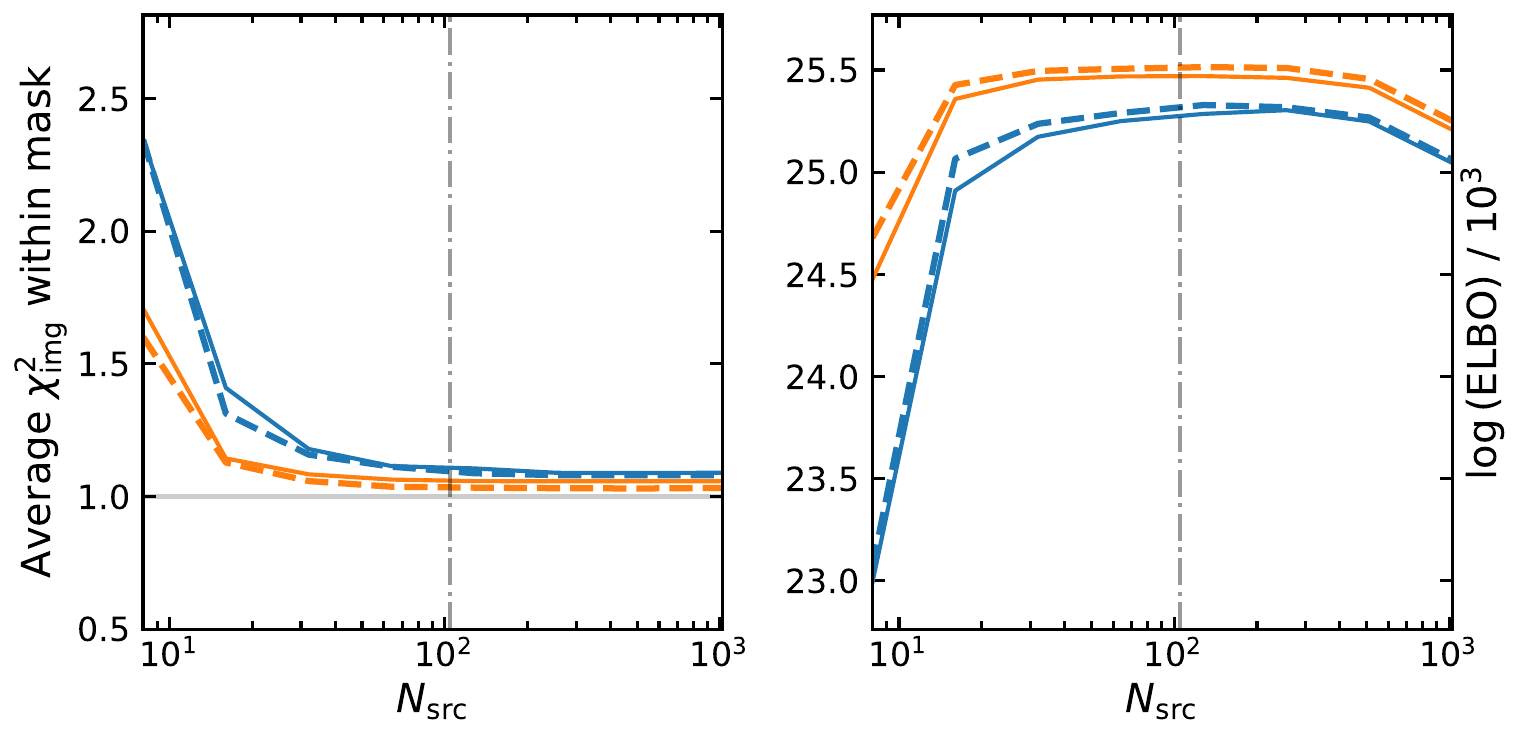} \\
    \caption{
    The evolution of $\chi_{\rm img}^2$ within the mask (left panels) and the model $\log(\mathrm{ELBO})$ (right panels). We show these diagnostics for Mock 1, Mock 2, Mock 3, and SDSS J0946+1006 (from top to bottom). Dashed lines indicate models with a substructure, while models without one are shown with solid lines. The vertical dot-dashed line marks the rule-of-thumb maximum $N_{\rm src}$ along each dimension to avoid strong prior interpolation. For each $N_{\rm src}$ we show the values for the chain with the highest $\log(\mathrm{ELBO})$.
    }
    \label{fig:Chi2_ELBO}
\end{figure}

\begin{figure}
    \centering
    \includegraphics[width=\hsize]{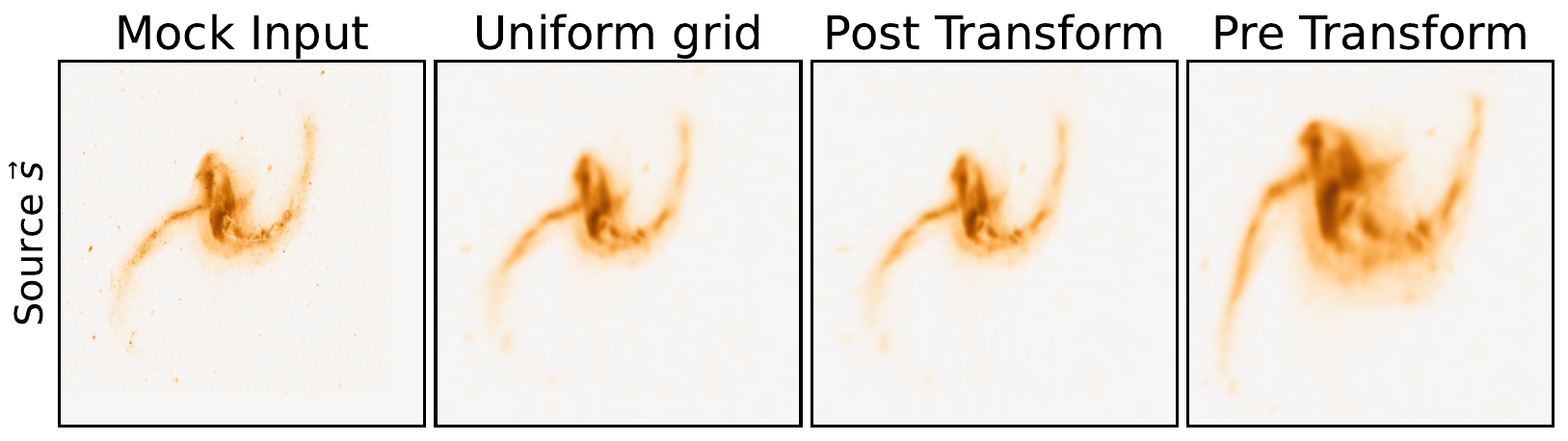} \\
    \includegraphics[width=\hsize]{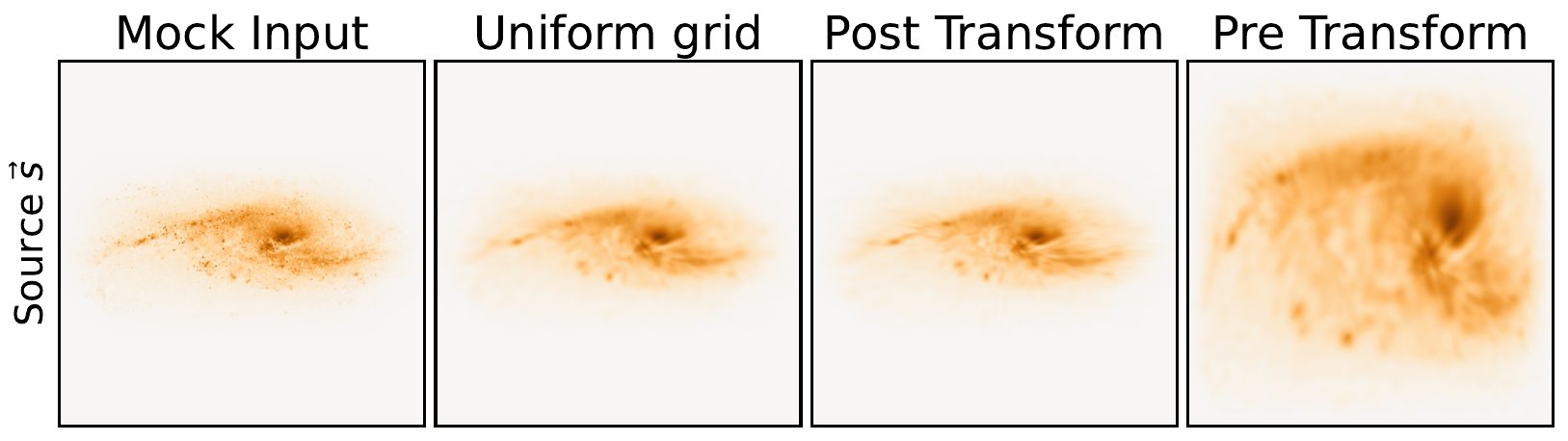} \\
    \includegraphics[width=\hsize]{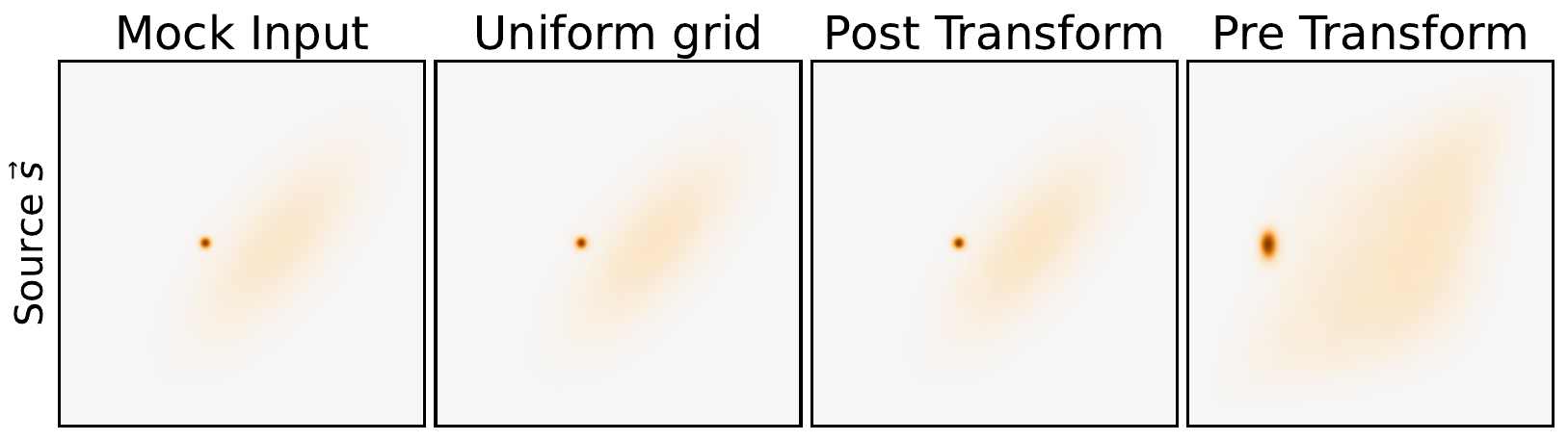} \\
    \includegraphics[width=\hsize]{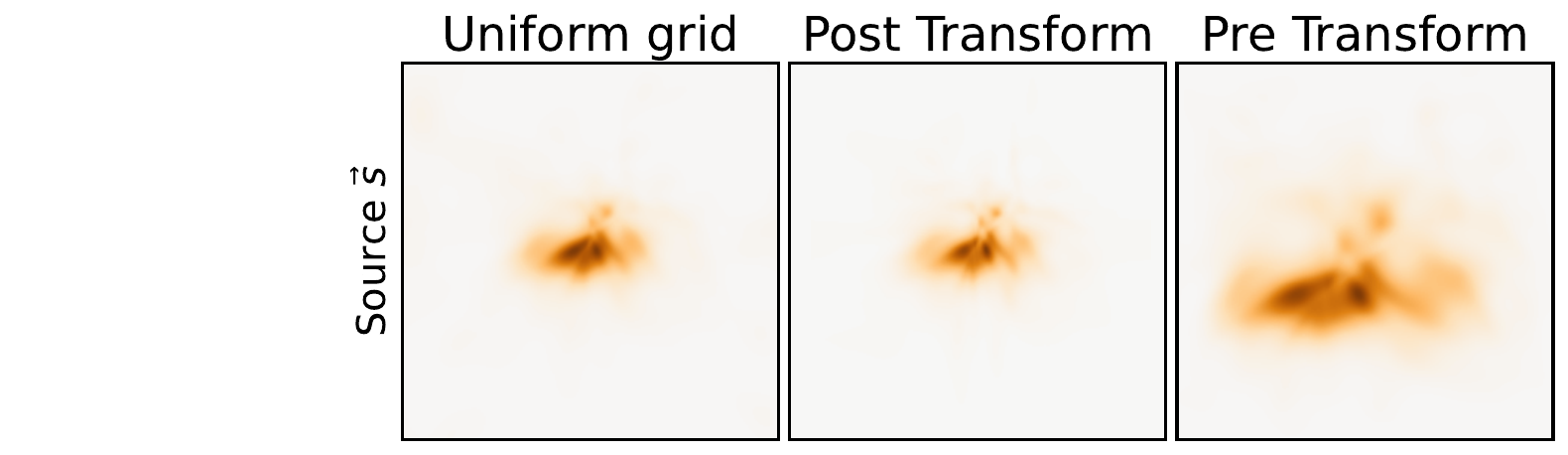} \\
    \caption{From left to right, for the three mocks we show the mock input source, the reconstructed source on a uniform grid, the reconstructed source on an RTU grid after the transformation is applied, and the reconstructed source on the RTU grid before the transformation is applied. For SDSS J0946+1006 no input source is available, so we show only the three reconstructed sources. From top to bottom we show Mock 1, Mock 2, Mock 3, and SDSS J0946+1006. The last column shows the reconstructed source before the transformation is applied, and therefore where most of the resolution for the reconstructed source is placed. The areas within the source that are imaged four times appear four times larger. The shown reconstructions have a pixel number of $N_{\rm src} = 1024$. Shown are the reconstructions from the chains with the highest ELBO.}
    \label{fig:SourcesDetail}
\end{figure}

\begin{figure*}
    \centering
    \includegraphics[width=0.32\textwidth]{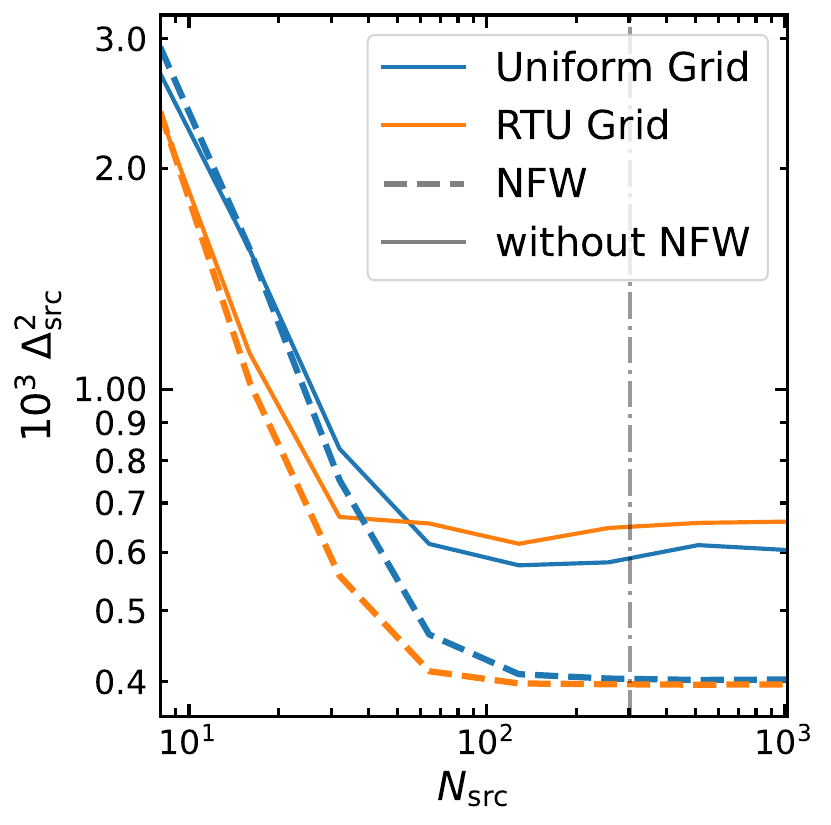} 
    \includegraphics[width=0.32\textwidth]{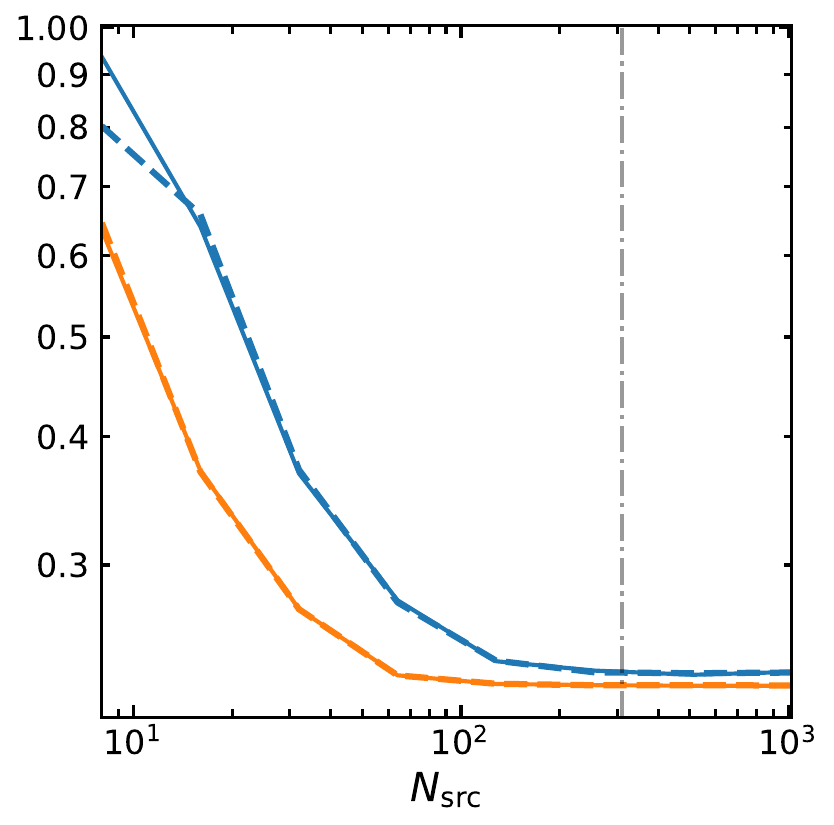} 
    \includegraphics[width=0.32\textwidth]{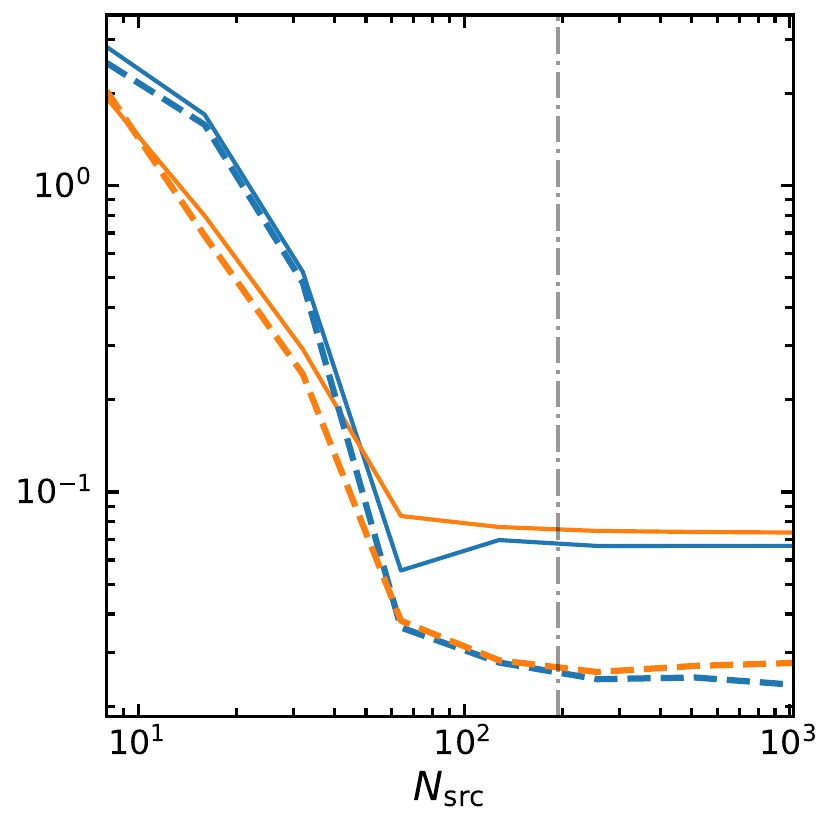}
    \caption{The evolution of $\Delta_{\rm src}^2$  for Mock 1 (left), Mock 2 (middle), and Mock 3 (right).}
    \label{fig:Chi2_src}
\end{figure*}

\section{Test cases}\label{sec:mock_generation} 
\subsection{Mock generation} 

We generate three mocks (shown in Figure \ref{fig:Mocks}) to test the RTU grid approach outlined above. All mocks assume JWST NIRCAM quality in the F150W filter, i.e. they have a pixel width $\Delta x = 0.031 \arcsec$ and a Point Spread Function with a Full Width Half Maximum of $\text{FWHM} = 2\Delta x$. All mocks are generated to have a size of $N_{\rm img}=200$ pixels along each side. To ensure that the sources show significant structure, we use high-resolution HST images of star-forming galaxies from the NGC catalog: NGC2623 for Mock 1 and NGC4536 for Mock 2. We also include a smooth source composed of two Gaussians for Mock 3 to provide a comparison. The sources are shown in Figure \ref{fig:SourcesDetail}. We assume Gaussian noise with an average signal-to-noise ratio (SNR) over the arcs of 12-15 (Maximum values range from 40 to 60). We employ a high super-sampling factor of $s_{\rm super} = 10$ when generating our mock images, to ensure that most of the source structure is accurately accounted for in our mocks. The three mock observations have deflectors that follow EPL+Shear models, which are a common choice for galaxy-galaxy lens models. We create these mocks to test different challenges of source reconstruction: 
i) Mock 1 includes a massive substructure with an NFW mass profile, chosen such that it is clearly detectable given the Signal-to-Noise and resolution of the mock. Its parameters resemble those inferred for the substructure in  SDSS J0946+1006 (the 'Jackpot' lens system) by \citet{Enzi25}. Mock 1 has a background source with a high dynamic range and complex structure. ii) Mock 2 contains no substructure. The source has complex structure and extends over a large image area. Using this mock we can probe whether we can exclude the presence of a substructure. iii) Mock 3 has a smooth source consisting of two Gaussians, one of which is much smaller and brighter than the other. With the smaller brighter source features near the caustics, we can probe if small scale features near the caustic benefit from the redistributed resolution of the adaptive grid. This mock also has a small mass NFW substructure.

As described in the previous section, we use masks to guide the RTU grid. Our masks include those pixels in which the SNR of the arcs is above some threshold.\footnote{After convolution with a Gaussian kernel with a standard deviation of 15 image pixels. The threshold is chosen to fully cover the arcs for each mock individually.} This produces the masks shown in Figure \ref{fig:Mocks}. We show (the ground truth parameters for the mocks and) our priors in Tables \ref{tab:SVI_Results_mock1}, \ref{tab:SVI_Results_mock2}, \ref{tab:SVI_Results_mock3}  and \ref{tab:jackpot} respectively. More detail on our model assumptions are given in Appendix \ref{sec:StrongGravitationalLensing}.

\subsection{SDSS J0946+1006}

We further test the RTU grid with ACS/F814W data of SDSS J0946+1006. This data has an exposure time of $\tau_{\rm exp} = 2096$ s. The lens system is part of the well-studied SLACS sample
of lenses \citep[][]{2008ApJ...677.1046G} and we use the data drizzled to 0.05 arcsec per pixel by \citet{2014MNRAS.443..969C}. We consider a $104\times104$ pixels cutout.  The
lens system shows a lens galaxy at redshift $z = 0.222$ and (among other sources) a source at $z_{s} = 0.609$. We will not consider other sources for the test of the RTU grid.

\section{Posterior and Analysis}
\label{sec:Analysis}

Following the description in \citet[][]{Enzi25}, we perform Stochastic Variational Inference (SVI) on all three mocks, and find a Gaussian approximation to the true posterior. The free parameters of these Gaussian distributions are their location and scale. Unlike \citet[][]{Enzi25}, we will not perform full Hamiltonian Monte Carlo (HMC) sampling, because the mock data is simple enough that the SVI provides a good enough posterior fit as well as Evidence Lower Bound (ELBO) for the purpose of this paper.\footnote{Error bars tend to be underestimated, but the underlying model is perfectly known so that fitting will be more robust.}  We consider the following source pixel numbers along each dimension $N_{\rm src} \in [8,16,32,64,128,256,512,1024]$. We run 10 SVI chains with 15000 steps for each test case. We always consider two models, one with and one without an included NFW substructure. We assume that the approximate location of the substructure is known and therefore do not perform a blind search for substructures. Whenever possible, we initialize the SVI from the parameters of the same model with the previously lower $N_{\rm src}$ value. The smallest values of  $N_{\rm src}$ are initialized using {\sc Numpyro}'s \texttt{init\_to\_median} function. 

We consider the following metrics as a function of $N_{\rm src}$ along each dimension to measure the quality of the two source reconstruction methods: i) The average normalized residuals $\chi_{\rm img}^2$ within the arc masks on the image plane. An average $\chi^2_{\rm img} \approx 1$ is expected for a good fit to the data. $\chi^2_{\rm img} \gg 1$ indicates a poor fit of the model to the data. $\chi^2_{\rm img} \ll 1$ indicates overfitting.

ii) The average difference between the reconstructed source and the true source after an overall spatial shift and scaling, $\Delta^2_{\rm src}$. This metric determines the quality of the reconstructed source when removing the effects of mass-sheet-like degeneracies. We calculate this by optimizing for the positional shift and scaling that provides the minimal $\Delta_{\rm src}^2$.\footnote{We use the ADABelief optimizer and bilinear interpolation to the grid of the original source.} We only apply this test for the mock data where the true source is available.

iii) The ELBO which is optimized during SVI and its difference between models with and without a substructure, which provide an indicator for how the adaptive grid affects the sensitivity to dark substructures within the lens. It is, however, important to note that SVI ELBOs are lower bounds on the Bayesian Evidence and the true marginal likelihood can vary between models.

\section{Results}
 \label{sec:Results}

The RTU grid achieves comparable fit quality to the uniform grid with a factor of $\simeq 2$ fewer pixels per dimension, i.e. $\simeq 4$ in total pixels. This holds across all three mocks and the real data. In the remainder of this section, we will substantiate this by considering the residuals, $\chi^2_{\rm img}$, $\Delta_{\rm src}^2$ and the ELBO. The results of our analyses are summarized in Tables \ref{tab:SVI_Results_mock1}, \ref{tab:SVI_Results_mock2},  \ref{tab:SVI_Results_mock3}, and \ref{tab:jackpot}. 

 SVI tends to be mode-seeking rather than mode-covering \citep[see e.g.][]{2016arXiv160202311L} and therefore error bars tend to be underestimated. By combining multiple independent SVI chains, we obtain more conservative estimates. This combination is performed in the unconstrained parameter space used internally by {\sc Numpyro}'s SVI implementation. We moment-match the Gaussian approximations from the independent SVI chains in this space, and then transform the resulting samples back to the constrained parameter space to compute the reported quantiles.
 We report the resulting constraints in Tables \ref{tab:SVI_Results_mock1}, \ref{tab:SVI_Results_mock2},  \ref{tab:SVI_Results_mock3}, and \ref{tab:jackpot} for $N_{\rm src} = 64$. A more rigorous treatment would require an HMC or equivalent sampling approach which is beyond the scope of this work. The RTU grid provides constraints that are comparable to and sometimes better than the uniform grid. Considering the inferred parameters and their uncertainty from multiple SVI chains, the recovered values are consistent with the ground-truth parameters within $2\sigma$ for $N_{\rm src} = 64$ along each dimension. The RTU grid tends to provide larger uncertainties.
 
In the case of the real SDSS J0946+1006 data, we find that the inferred parameters do not fully agree with those reported in \citet[][]{Enzi25} within $2\sigma$. This is expected for multiple reasons: (i) we assume a simpler model in this work, which does not include multipoles and/or the arc of the second source; (ii) SVI posteriors are not as rigorous as full HMC samples; and (iii) we did not include any light from the substructure, which can be important for reconstructing the substructure parameters \citep[see e.g.][]{2025ApJ...991L..53H}.

Figures \ref{fig:ReconstructionMock1}, \ref{fig:ReconstructionMock2},  \ref{fig:ReconstructionMock3},  and \ref{fig:ReconstructionJackpot} show the reconstructed sources and normalized residuals as a function of $N_{\rm src}$ along each dimension. We find that the $N_{\rm src}$ required for noise-level residuals using the RTU grid is reduced by a factor of $\simeq 2$ for the three mocks we consider. This is in agreement with our expectation from the number of images present in the (mock) data, when the image multiplicity is the dominant contribution to the ray density.

Figure \ref{fig:Chi2_ELBO} shows the evolution of $\chi^2_{\rm img}$ as a function of $N_{\rm src}$. The $\chi^2_{\rm img}$ indicates image-plane fit quality and both source grid approaches converge towards the limit of $\chi^2_{\rm img} \approx 1$. This value indicates a good fit and is achieved for a factor of two fewer pixels per dimension ($\simeq$4 in total) using the RTU approach. Mild overfitting (indicated by $\chi^2_{\rm img} \lesssim 1.0$) happens in the case of Mock 1 and Mock 2, which may arise from a residual background that has not been perfectly subtracted from the input sources during mock generation. We note that the $\chi^2_{\rm img}$ can show a non-monotonic behaviour that is not necessarily unexpected. The SVI does not optimize the $\chi^2_{\rm img}$ but the ELBO in order to provide a posterior estimate. We observed no overfitting for  SDSS J0946+1006, which is not surprising given the absence of the additional model complexity that previous studies of this object considered \citep[e.g. multipoles in the lens mass distribution][]{Enzi25,2025ApJ...991L..53H,2025MNRAS.543..540T}.

Figure \ref{fig:Chi2_ELBO} further shows the evolution of ELBOs as a function of $N_{\rm src}$. In all cases, we find that the ELBO first increases and then plateaus, with a slight downturn in the case of the real SDSS J0946+1006 data. The RTU grid plateaus earlier than the uniform grid, with half as many pixels per dimension, $N_{\rm src}$. 
For the real data of SDSS J0946+1006 the ELBO reaches its maximum around the pixel number corresponding to the previously described rule of thumb. 

Having established the convergence behaviour of the two grids, we next use the ELBO to compare lens models with and without substructure. Since the ELBO is a lower bound on the evidence, it can provide some indication for model preference. We find for Mock 1 a significant increase in $\Delta \log$(ELBO) $> 4000$ when the substructure is included, suggesting a strong preference for the substructure model under this variational approximation. In the case of Mock 2, the model with a substructure shows around the same or a lower ELBO than the model without a substructure. While a substructure might not be ruled out, its mass would be required to be negligibly small. For Mock 3, we find that the ELBO increases with the inclusion of the substructure, albeit not as significantly as for Mock 1. This is due to the smooth source, which renders this mock less sensitive to the presence of the included substructure. With a difference of $\Delta \log$(ELBO) $\approx35$, the ELBO comparison suggests a preference for the model with an NFW substructure for SDSS J0946+1006 when the RTU grid is employed.

Finally, we test whether the reconstructed source itself is accurate. To this end, we consider the difference with respect to the original source, $\Delta^2_{\rm src}$ while ignoring translation and scaling of the source, which could be explained by a mass-sheet transformation. Figure \ref{fig:Chi2_src} shows the evolution of $\Delta^2_{\rm src}$ with $N_{\rm src}$. As expected, when the original mock contains a substructure, the source reconstruction is more accurate for models that include this substructure. For these mocks, the reconstruction without the NFW substructure is worse in the RTU reconstruction. Overall, when the model correctly accounts for the presence or absence of a substructure, the RTU grid achieves more accurate source reconstructions with smaller $N_{\rm src}$.

\section{Discussion} 
\label{sec:Discussion} 

\subsection{Limitations}

If the ray density on the source plane is already uniform in the marginals, the proposed RTU grid approaches the uniform grid, because the ray-guided transformation becomes approximately the identity.
This could happen when the lens mapping produces a complicated caustic
network, e.g. in microlensing. The RTU approach is then likely not going to improve the reconstruction significantly. Based on the histograms in Figure \ref{fig:Mocks}, we do not expect this to be a major concern for the galaxy-galaxy lensing systems considered here. This assumption might not hold for other cases, such as cluster scale lens systems. 

Another limitation is the dependence on the coordinate axes used to construct
the transformations. Since the RTU grid is built from marginal ray-density
distributions, a strongly tilted or highly elliptical ray-density pattern may
not be captured optimally in the original coordinate system. This may matter for very elliptical cluster lens systems, it does not appear to be a major factor in the galaxy-galaxy mock lenses that we generated for this work. If required, the coordinate system can be rotated before applying the transformation. 

\subsection{Interpretation}

Finally, the transformation affects how the resulting power spectrum should be interpreted. Similar to cells in methods based on Delaunay or Voronoi tessellation, some pixels appear more stretched than others. This affects the correlation structure in a similar way to how stretched cells affect gradient and curvature regularization in mesh based methods. Derivatives far away from the caustics will typically be evaluated over larger physical source-plane scales than derivatives close to the caustics. For details on how this stretching affects the correlation structure, we refer to Appendix \ref{sec:GPafterT}.

\subsection{Extension to weighted RTU grid}

While we did not employ this extension in this paper, it is also possible to weight the rays by importance when constructing the eCDFs, resulting in a weighted RTU grid. This can be achieved e.g. by choosing weights according to the image pixel brightness, gradient, or the SNR. Such weighting would place emphasis on the most important source structure and is similar to the approach used by \citet[][]{2021JOSS....6.2825N}. Such weights also make the grid depend not only on the lens mapping but also on the observed image properties. While we do not discuss such weighting schemes in detail, the weighted RTU is already implemented and made available alongside this paper.

\section{Conclusions}

\label{sec:Conclusion}

The prior assumptions of source reconstruction remain  one of the main uncertainties in lens modelling and can be a major source of bias in inferred lens models. While  standard regularization methods have been highly successful, source  reconstruction approaches often involve a trade-off between realism, flexibility, auto-differentiability, and computational efficiency simultaneously. The RTU grid method presented here addresses this trade-off by applying a transform such that the resulting source pixels are uniformly covered by rays traced back to the source plane.

We find that the RTU grid leads to higher ELBOs and achieves a comparable image fit quality to the uniform grid for $N_{\rm src}$ that are smaller by a factor of roughly $2$ in each dimension.
When the mass model is accurate, the RTU reaches comparable $\chi^2_{\rm img}$ and typically a lower $\Delta^2_{\rm src}$ than the uniform grid. At the same time, the ELBO difference between models with and without a substructure is not significantly affected by the RTU grid. This indicates that the improved efficiency does not drive the inferred preferences for substructure.

With the arrival of large samples of lensing data from Euclid \citep[see e.g.][]{2025arXiv250315326E} and high-resolution, high-SNR follow-up data, e.g. from JWST, robust and flexible source modelling approaches will be increasingly important. An implementation of the RTU grid approach for source reconstruction compatible with {\sc Herculens} \citep[][]{2022A&A...668A.155G} is made available alongside this paper. \footnote{\href{https://github.com/CKrawczyk/RTU-grid-herculens}{https://github.com/CKrawczyk/RTU-grid-herculens} \label{footnote:URL}}

\section*{Acknowledgements}

We thank James Nightingale for helpful discussion and feedback. This project has received
funding from the European Research Council (ERC) under the European Union’s Horizon 2020 research and innovation programme
(LensEra: grant agreement No 945536). TEC is funded by the Royal
Society through a University Research Fellowship. Numerical computations were done on the Sciama High Performance Compute
(HPC) cluster which is supported by the ICG, SEPNet and the University of Portsmouth. 
For the purpose of open access, the authors have applied a Creative Commons Attribution (CC BY) licence to any Author Accepted Manuscript version arising.

\section*{Data Availability}

Supporting research data are available on reasonable request from the corresponding author.



\bibliographystyle{mnras}
\bibliography{example} 




\appendix

\section{Parametric models for mass and lens light}
\label{sec:StrongGravitationalLensing}

\subsection{Elliptical Power Law with Shear}
We assume main deflectors that follow an elliptical power-law (EPL) with additional external Shear. The convergence distribution of this power law mass profile is given as:
    \begin{equation}
        \kappa_{\rm EPL}(\rho,\gamma,q) = \frac{3-\gamma}{2} \left(\frac{\theta_E}{\rho}\right)^{\gamma-1} \,,
    \end{equation}
with the Einstein radius $\theta_E$, the logarithmic slope $\gamma$, and the elliptical radius $\rho^2 = (x-x_0)^2q + (y-y_0)^2/q $. The position angle, the minor-to-major axis ratio $q$ and center $x_0,y_0$ are inferred during lens modelling. We parametrize these via their ellipticities \citep[see e.g.][]{Enzi25}. In addition to the EPL we include external Shear characterized by $\vec \Gamma = (\Gamma_1,\Gamma_2)$ as defined in \citet{Enzi25}.

\subsection{NFW substructure}
Two of the mocks we create include substructures in the form of an NFW density profile \citep[][]{1997ApJ...490..493N}. The three dimensional NFW profile is characterized by its characteristic density $\rho_s$ and scale radius $r_s$:
\begin{equation}
\rho^{\rm 3D}_{\rm NFW}(r) =   \frac{\rho_s  }{ 
\frac{r}{r_s} (1+ \frac{r}{r_s})^2 }\,.
\end{equation}
In projection, this gives rise to a convergence of the following form \cite[see e.g.][]{2001astro.ph..2341K}:
\begin{equation}
\kappa_{\rm NFW}(r) =  2 \kappa_s\Big( 1.0 - F(r/r_s) \Big) \Big/ \Big( (r/r_s)^2  - 1\Big) \,,
\end{equation}
with $r^2= x^2+y^2$ , $r_s$ being the scale radius of the NFW profile, and $\kappa_s$ determining the amplitude of the convergence. $F(x) = \tan^{-1} \big(\sqrt{x^2-1}\big) / \sqrt{x^2-1}$ for $x>1$ and $\tanh^{-1} \big(\sqrt{1-x^2}\big) / \sqrt{1-x^2}$ for $0<x\leq1$.
\subsection{Multi Gaussian Expansion}
We assume that the lens light is well represented by a sum of Gaussian components, which has been shown to provide a good fit to the lens light in real observations \citep[][]{2024MNRAS.532.2441H}:
\begin{equation}
s_{\rm lens}(\vec x) = \sum_{n=1}^{N_{\rm gauss}} \mathcal{G}(\rho_n, \sigma_n) \,.
\end{equation}
 Each Gaussian component is determined as:
\begin{equation}
\mathcal{G}(\rho_n, \sigma_n) = \frac{A_n}{2\pi } \exp \Big( - \frac{1}{2} \rho_n^2 / \sigma_n^2 \Big) \,, 
\end{equation}
where we allow the amplitudes $A_n$, the minor-to-major axis ratios $q_n$ that appear in the radius $\rho_n^2=(x-x_n)^2q_n+(y-y_n)^2/q_n$,  centers, and position angles to be free parameters for each component individually. Throughout this paper we will always assume $N_{\rm gauss} = 5$. To alleviate a strongly multi-modal probability distribution, but allowing for enough freedom in the lens light reconstruction with the fewest Gaussian components, we draw all standard deviations from the same range $\sigma_n\in [0.1, 1.0]$ with a Log-Uniform prior. We then sort these $\sigma$ values and draw the other lens light parameters accordingly. While the resulting distribution can still be multi-modal, secondary modes will become less likely through this process.
%

\renewcommand{\arraystretch}{1.3}
\begin{table*}

	\centering
	\caption{ This table presents the truth value, our choice of prior distribution, and the posterior constraints of the main model parameters for Mock 1. We refer to uniform distributions as $\mathcal{U}$, normal distributions as $\mathcal{G}$, halfnormal distributions as $\mathcal{G}_+$, and  the linear distribution on a finite interval as described in appendix of \citet[][]{Enzi25} as $\mathcal{T}$. Here we colored the cells of the model matching the ground truth with colors that indicate the quality of the fit. Red indicates a reconstructed value that is too low, blue indicates a reconstructed value that is too high. The cells background is lightgray when the reconstructed value is within 1 standard deviation of the reconstruction. The $\Delta \log$ ELBO values are reported relative to the uniform-grid smooth model in each table.
    } 
	\label{tab:SVI_Results_mock1}
 \small

\begin{tabular}{lccccccc} 
\hline
\textbf{Parameter}  & \textbf{Truth} & \textbf{Prior}  & \multicolumn{2}{c}{\textbf{Uniform} (Median $\pm 2\sigma$)}   & \multicolumn{2}{c}{\textbf{Transformed Uniform} (Median $\pm 2\sigma$)}  \\
& & &Smooth & NFW & Smooth  & NFW \\
\hline

\hline

$\log_{10} \kappa_s$ & 0.3  & ${\mathcal{U}(-10,4)}$  &-&\cellcolor[RGB]{247,246,246}  $-1.54^{+5.41}_{-6.01}$ &-&\cellcolor[RGB]{247,246,246}  $-2.18^{+6.02}_{-6.50}$ \\
$\log_{10} r_s / \arcsec$ & -1.42 &  ${\mathcal{U}(-4,4)}$  &-&\cellcolor[RGB]{246,246,246}  $-1.30^{+1.40}_{-1.38}$ &-&\cellcolor[RGB]{246,246,246}  $-0.74^{+2.79}_{-2.67}$ \\
$x_{\rm sub} [\arcsec] + 1.6$ & 0.0 &  $\mathcal{G}(0.0,0.05)$  &-&\cellcolor[RGB]{247,246,246}  $-0.00006^{+0.05279}_{-0.05279}$ &-&\cellcolor[RGB]{247,246,246}  $-0.06^{+0.33}_{-0.33}$ \\
$y_{\rm sub} [\arcsec]$ & 0.0 & $\mathcal{G}(0.0,0.05)$  &-&\cellcolor[RGB]{246,246,246}  $0.001^{+0.053}_{-0.053}$ &-&\cellcolor[RGB]{247,246,246}  $-0.07^{+0.39}_{-0.39}$ \\
\hline
 $\theta_E [\arcsec]$ & 1.5 &  ${\mathcal{U}(-10,15)}$ & $1.502^{+0.004}_{-0.004}$ &\cellcolor[RGB]{252,221,202}  $1.499^{+0.001}_{-0.001}$ & $1.5019^{+0.0002}_{-0.0002}$ &\cellcolor[RGB]{246,246,246}  $1.50^{+0.08}_{-0.05}$ \\
$\gamma \in [1.0,3.0]$ &  2.1&  ${\mathcal{G}(2.0,0.2)}$ & $2.18^{+0.10}_{-0.11}$ &\cellcolor[RGB]{228,238,243}  $2.17^{+0.11}_{-0.11}$ & $2.153^{+0.001}_{-0.001}$ &\cellcolor[RGB]{246,246,246}  $2.23^{+0.44}_{-0.56}$ \\
$x [\arcsec]$ &0.01 &  ${\mathcal{G}(0,0.5)}$ & $0.013^{+0.003}_{-0.003}$ &\cellcolor[RGB]{247,246,246}  $0.009^{+0.003}_{-0.003}$ & $0.0145^{+0.0004}_{-0.0004}$ &\cellcolor[RGB]{247,246,246}  $-0.02^{+0.16}_{-0.16}$ \\
$y [\arcsec]$ &0.0 &  ${\mathcal{G}(0,0.5)}$ &  $-0.010^{+0.004}_{-0.004}$ &\cellcolor[RGB]{247,246,246}  $-0.005^{+0.011}_{-0.011}$ & $-0.0090^{+0.0003}_{-0.0003}$ &\cellcolor[RGB]{246,246,246}  $0.10^{+0.30}_{-0.30}$ \\
$e_x$ & 0.1&  ${\mathcal{G}(0.0,0.1)}$ & $0.13^{+0.01}_{-0.01}$ &\cellcolor[RGB]{246,246,246}  $0.11^{+0.03}_{-0.03}$ & $0.1168^{+0.0006}_{-0.0006}$ &\cellcolor[RGB]{247,246,246}  $-0.03^{+0.37}_{-0.37}$ \\
$e_y$ & -0.05&  ${\mathcal{G}(0.0,0.1)}$ & $-0.040^{+0.005}_{-0.005}$ &\cellcolor[RGB]{246,246,246}  $-0.049^{+0.007}_{-0.007}$ & $-0.0371^{+0.0002}_{-0.0002}$ &\cellcolor[RGB]{246,246,246}  $0.07^{+0.33}_{-0.33}$ \\
$\Gamma_1$ & -0.03 &  ${\mathcal{U}(-0.2,0.2)}$ & $-0.04^{+0.01}_{-0.01}$ &\cellcolor[RGB]{250,229,217}  $-0.04^{+0.01}_{-0.01}$ & $-0.0378^{+0.0002}_{-0.0002}$ &\cellcolor[RGB]{247,246,246}  $-0.08^{+0.18}_{-0.10}$ \\
$\Gamma_2$ & 0.02 &  ${\mathcal{U}(-0.2,0.2)}$ & $0.028^{+0.004}_{-0.004}$ &\cellcolor[RGB]{243,245,246}  $0.024^{+0.008}_{-0.009}$ & $0.0270^{+0.0001}_{-0.0001}$ &\cellcolor[RGB]{246,246,246}  $0.04^{+0.04}_{-0.04}$ \\
\hline
$n_{\rm src}$ & - &  ${\mathcal{T}(-1,0.1,10.0)}$ & $0.261^{+0.008}_{-0.007}$ & $0.25^{+0.01}_{-0.01}$ & $0.31^{+0.02}_{-0.01}$ & $0.16^{+1.73}_{-0.06}$ \\
$\zeta_{\rm src}$ & - &  $10^{\mathcal{G}(2.1,1.1)}$ & $17.70^{+2.07}_{-1.85}$ & $18.98^{+1.87}_{-1.70}$ & $26.19^{+2.11}_{-1.95}$ & $20.31^{+19.12}_{-9.85}$ \\
$\sigma_{\rm src}$ & - &  $10^{\mathcal{U}(-5,2)}$  & $0.139^{+0.004}_{-0.004}$ & $0.142^{+0.007}_{-0.006}$ & $0.202^{+0.007}_{-0.007}$ & $0.22^{+0.07}_{-0.06}$ \\
$\alpha_{\rm src}$ & - &  $10^{\mathcal{U}(-1,1)}$ & $1.66^{+0.02}_{-0.02}$ & $1.67^{+0.02}_{-0.02}$ & $1.93^{+0.01}_{-0.01}$ & $1.95^{+0.26}_{-0.24}$ \\
\hline

$\log$ ELBO&&&109986.31&115546.55&112578.15&117153.03\\
$\Delta\log$  ELBO &&&0.00&5560.24&2591.84&7166.72\\
\hline
\end{tabular}

\end{table*}


\begin{table*}

	\centering
	\caption{ Same as Table \ref{tab:SVI_Results_mock1}, but for Mock 2.}
    \label{tab:SVI_Results_mock2}
 \small

\begin{tabular}{lccccccc} 
\hline
\textbf{Parameter}  & \textbf{Truth} & \textbf{Prior}  & \multicolumn{2}{c}{\textbf{Uniform} (Median $\pm 2\sigma$)}   & \multicolumn{2}{c}{\textbf{Transformed Uniform} (Median $\pm 2\sigma$)}  \\
& & &Smooth & NFW & Smooth  & NFW \\
\hline

\hline

$\log_{10} \kappa_s$ & -  & ${\mathcal{U}(-10,4)}$  &-& $-4.02^{+5.36}_{-5.26}$ &-& $-4.37^{+5.89}_{-5.44}$ \\
$\log_{10} r_s / \arcsec$ & - &  ${\mathcal{U}(-4,4)}$  &-& $-0.83^{+3.05}_{-2.81}$ &-& $-0.82^{+3.08}_{-2.83}$ \\
$x_{\rm sub} [\arcsec] + 1.6$ & - &  $\mathcal{G}(0.0,0.05)$  &-& $-0.10^{+0.46}_{-0.46}$ &-& $-0.05^{+0.78}_{-0.78}$ \\
$y_{\rm sub} [\arcsec]$ & - & $\mathcal{G}(0.0,0.05)$  &-& $-0.04^{+0.52}_{-0.52}$ &-& $0.02^{+0.41}_{-0.41}$ \\
\hline
 $\theta_E [\arcsec]$ & 1.5 &  ${\mathcal{U}(-10,15)}$ &\cellcolor[RGB]{246,246,246}  $1.54^{+0.10}_{-0.07}$ & $1.51^{+0.06}_{-0.04}$ &\cellcolor[RGB]{246,246,246}  $1.51^{+0.05}_{-0.04}$ & $1.50^{+0.03}_{-0.03}$ \\
$\gamma \in [1.0,3.0]$ &  2.1&  ${\mathcal{G}(2.0,0.2)}$ &\cellcolor[RGB]{247,246,246}  $2.06^{+0.16}_{-0.17}$ & $2.10^{+0.11}_{-0.11}$ &\cellcolor[RGB]{247,246,246}  $2.08^{+0.07}_{-0.07}$ & $2.26^{+0.52}_{-0.72}$ \\
$x [\arcsec]$ &0.01 &  ${\mathcal{G}(0,0.5)}$ &\cellcolor[RGB]{246,246,246}  $0.04^{+0.30}_{-0.30}$ & $0.002^{+0.130}_{-0.130}$ &\cellcolor[RGB]{247,246,246}  $-0.001^{+0.068}_{-0.068}$ & $0.05^{+0.09}_{-0.09}$ \\
$y [\arcsec]$ &0.0 &  ${\mathcal{G}(0,0.5)}$ & \cellcolor[RGB]{246,246,246}  $0.005^{+0.238}_{-0.238}$ & $0.04^{+0.22}_{-0.22}$ &\cellcolor[RGB]{247,246,246}  $-0.005^{+0.028}_{-0.028}$ & $0.03^{+0.38}_{-0.38}$ \\
$e_x$ & 0.1&  ${\mathcal{G}(0.0,0.1)}$ &\cellcolor[RGB]{247,246,246}  $0.05^{+0.16}_{-0.16}$ & $0.09^{+0.08}_{-0.08}$ &\cellcolor[RGB]{247,246,246}  $0.09^{+0.06}_{-0.06}$ & $0.05^{+0.29}_{-0.29}$ \\
$e_y$ & -0.05&  ${\mathcal{G}(0.0,0.1)}$ &\cellcolor[RGB]{246,246,246}  $-0.01^{+0.19}_{-0.19}$ & $-0.04^{+0.06}_{-0.06}$ &\cellcolor[RGB]{246,246,246}  $-0.04^{+0.09}_{-0.09}$ & $0.05^{+0.27}_{-0.27}$ \\
$\Gamma_1$ & -0.03 &  ${\mathcal{U}(-0.2,0.2)}$ &\cellcolor[RGB]{246,246,246}  $-0.01^{+0.08}_{-0.08}$ & $-0.02^{+0.05}_{-0.05}$ &\cellcolor[RGB]{246,246,246}  $-0.03^{+0.03}_{-0.03}$ & $-0.02^{+0.06}_{-0.06}$ \\
$\Gamma_2$ & 0.02 &  ${\mathcal{U}(-0.2,0.2)}$ &\cellcolor[RGB]{246,246,246}  $0.02^{+0.07}_{-0.08}$ & $0.02^{+0.03}_{-0.03}$ &\cellcolor[RGB]{246,246,246}  $0.02^{+0.02}_{-0.02}$ & $0.02^{+0.05}_{-0.06}$ \\
\hline
$n_{\rm src}$ & - &  ${\mathcal{T}(-1,0.1,10.0)}$ & $0.25^{+0.24}_{-0.09}$ & $0.24^{+0.31}_{-0.10}$ & $0.38^{+0.10}_{-0.07}$ & $0.23^{+0.44}_{-0.10}$ \\
$\zeta_{\rm src}$ & - &  $10^{\mathcal{G}(2.1,1.1)}$ & $30.92^{+4.19}_{-3.69}$ & $29.11^{+5.79}_{-4.83}$ & $32.64^{+2.48}_{-2.31}$ & $35.33^{+7.80}_{-6.39}$ \\
$\sigma_{\rm src}$ & - &  $10^{\mathcal{U}(-5,2)}$  & $0.109^{+0.009}_{-0.008}$ & $0.103^{+0.007}_{-0.006}$ & $0.16^{+0.01}_{-0.01}$ & $0.16^{+0.01}_{-0.01}$ \\
$\alpha_{\rm src}$ & - &  $10^{\mathcal{U}(-1,1)}$ & $1.59^{+0.06}_{-0.06}$ & $1.59^{+0.11}_{-0.11}$ & $1.814^{+0.009}_{-0.008}$ & $1.92^{+0.18}_{-0.17}$ \\
\hline

$\log$ ELBO&&&111847.13&111847.56&115536.24&115534.22\\
$\Delta\log$  ELBO &&&0.00&0.44&3689.11&3687.09\\
\hline
\end{tabular}

\end{table*}


\begin{table*}

	\centering
	\caption{ Same as Table \ref{tab:SVI_Results_mock1}, but for Mock 3.}
    \label{tab:SVI_Results_mock3}

\begin{tabular}{lccccccc} 
\hline
\textbf{Parameter}  & \textbf{Truth} & \textbf{Prior}  & \multicolumn{2}{c}{\textbf{Uniform} (Median $\pm 2\sigma$)}   & \multicolumn{2}{c}{\textbf{Transformed Uniform} (Median $\pm 2\sigma$)}  \\
& & &Smooth & NFW & Smooth  & NFW \\
\hline

\hline

$\log_{10} \kappa_s$ & -2.0  & ${\mathcal{U}(-10,4)}$  &-&\cellcolor[RGB]{247,246,246}  $-2.77^{+4.52}_{-4.52}$ &-&\cellcolor[RGB]{247,246,246}  $-2.63^{+4.37}_{-4.37}$ \\
$\log_{10} r_s / \arcsec$ & 0.0 &  ${\mathcal{U}(-4,4)}$  &-&\cellcolor[RGB]{247,246,246}  $-0.36^{+2.04}_{-2.03}$ &-&\cellcolor[RGB]{247,246,246}  $-0.17^{+1.75}_{-1.75}$ \\
$x_{\rm sub} [\arcsec] + 0.92$ & 0.0 &  $\mathcal{G}(0.0,0.05)$  &-&\cellcolor[RGB]{250,231,219}  $-0.05^{+0.09}_{-0.09}$ &-&\cellcolor[RGB]{246,246,246}  $0.04^{+0.16}_{-0.16}$ \\
$y_{\rm sub} [\arcsec] - 1.13$ & 0.0 & $\mathcal{G}(0.0,0.05)$  &-&\cellcolor[RGB]{247,246,246}  $-0.19^{+0.96}_{-0.96}$ &-&\cellcolor[RGB]{247,246,246}  $-0.06^{+0.15}_{-0.15}$ \\
\hline
 $\theta_E [\arcsec]$ & 1.5 &  ${\mathcal{U}(-10,15)}$ & $1.51^{+0.03}_{-0.04}$ &\cellcolor[RGB]{247,246,246}  $1.50^{+0.01}_{-0.01}$ & $1.506^{+0.005}_{-0.005}$ &\cellcolor[RGB]{247,246,246}  $1.49^{+0.05}_{-0.03}$ \\
$\gamma \in [1.0,3.0]$ &  2.1&  ${\mathcal{G}(2.0,0.2)}$ & $2.38^{+0.44}_{-0.72}$ &\cellcolor[RGB]{246,246,246}  $2.35^{+0.48}_{-0.77}$ & $2.49^{+0.43}_{-0.95}$ &\cellcolor[RGB]{210,229,240}  $2.44^{+0.33}_{-0.52}$ \\
$x [\arcsec]$ &0.01 &  ${\mathcal{G}(0,0.5)}$ & $0.05^{+0.12}_{-0.12}$ &\cellcolor[RGB]{246,246,246}  $0.02^{+0.05}_{-0.05}$ & $0.007^{+0.014}_{-0.014}$ &\cellcolor[RGB]{246,246,246}  $0.02^{+0.11}_{-0.11}$ \\
$y [\arcsec]$ &0.0 &  ${\mathcal{G}(0,0.5)}$ &  $-0.02^{+0.11}_{-0.11}$ &\cellcolor[RGB]{246,246,246}  $0.002^{+0.019}_{-0.019}$ & $0.007^{+0.027}_{-0.027}$ &\cellcolor[RGB]{247,246,246}  $-0.02^{+0.06}_{-0.06}$ \\
$e_x$ & 0.1&  ${\mathcal{G}(0.0,0.1)}$ & $0.04^{+0.31}_{-0.31}$ &\cellcolor[RGB]{246,246,246}  $0.14^{+0.13}_{-0.13}$ & $0.11^{+0.12}_{-0.12}$ &\cellcolor[RGB]{247,246,246}  $0.09^{+0.33}_{-0.33}$ \\
$e_y$ & 0.04&  ${\mathcal{G}(0.0,0.1)}$ & $-0.003^{+0.227}_{-0.227}$ &\cellcolor[RGB]{247,246,246}  $0.006^{+0.210}_{-0.210}$ & $-0.04^{+0.44}_{-0.44}$ &\cellcolor[RGB]{247,246,246}  $-0.002^{+0.224}_{-0.224}$ \\
$\Gamma_1$ & -0.003 &  ${\mathcal{U}(-0.2,0.2)}$ & $-0.03^{+0.06}_{-0.06}$ &\cellcolor[RGB]{247,246,246}  $-0.01^{+0.03}_{-0.03}$ & $-0.02^{+0.04}_{-0.04}$ &\cellcolor[RGB]{247,246,246}  $-0.03^{+0.05}_{-0.05}$ \\
$\Gamma_2$ & 0.002 &  ${\mathcal{U}(-0.2,0.2)}$ & $-0.005^{+0.045}_{-0.045}$ &\cellcolor[RGB]{247,246,246}  $-0.006^{+0.022}_{-0.022}$ & $-0.007^{+0.021}_{-0.021}$ &\cellcolor[RGB]{247,246,246}  $-0.009^{+0.029}_{-0.029}$ \\
\hline
$n_{\rm src}$ & - &  ${\mathcal{T}(-1,0.1,10.0)}$ & $0.17^{+0.76}_{-0.06}$ & $0.35^{+0.65}_{-0.18}$ & $0.50^{+2.89}_{-0.36}$ & $0.54^{+2.34}_{-0.39}$ \\
$\zeta_{\rm src}$ & - &  $10^{\mathcal{G}(2.1,1.1)}$ & $19.79^{+39.05}_{-13.13}$ & $28.67^{+20.56}_{-11.97}$ & $22.93^{+5.55}_{-4.47}$ & $25.12^{+7.86}_{-5.99}$ \\
$\sigma_{\rm src}$ & - &  $10^{\mathcal{U}(-5,2)}$  & $0.21^{+0.08}_{-0.06}$ & $0.20^{+0.02}_{-0.02}$ & $0.32^{+0.03}_{-0.03}$ & $0.35^{+0.06}_{-0.05}$ \\
$\alpha_{\rm src}$ & - &  $10^{\mathcal{U}(-1,1)}$ & $2.27^{+0.74}_{-0.60}$ & $2.74^{+0.59}_{-0.51}$ & $2.83^{+0.29}_{-0.28}$ & $2.94^{+0.66}_{-0.58}$ \\
\hline

$\log$ ELBO&&&117068.36&118147.90&118624.36&119519.49\\
$\Delta\log$  ELBO &&&0.00&1079.54&1556.00&2451.13\\
\hline
\end{tabular}

\end{table*}

\begin{table*}

	\centering
	\caption{ Same as Table \ref{tab:SVI_Results_mock1}, but for SDSS J0946+1006.}
    \label{tab:jackpot}

\begin{tabular}{lccccccc} 
\hline
\textbf{Parameter}  & \textbf{Truth} & \textbf{Prior}  & \multicolumn{2}{c}{\textbf{Uniform} (Median $\pm 2\sigma$)}   & \multicolumn{2}{c}{\textbf{Transformed Uniform} (Median $\pm 2\sigma$)}  \\
& & &Smooth & NFW & Smooth  & NFW \\
\hline

\hline

$\log_{10} \kappa_s$ &   & ${\mathcal{U}(-10,4)}$  &-& $-4.55^{+6.63}_{-5.42}$ &-& $-5.01^{+6.22}_{-4.97}$ \\
$\log_{10} r_s / \arcsec$ &   &  ${\mathcal{U}(-4,4)}$  &-& $-1.62^{+3.38}_{-2.33}$ &-& $-1.38^{+3.41}_{-2.55}$ \\
$x_{\rm sub} [\arcsec] + 0.68$ &   &  $\mathcal{G}(0.0,0.05)$  &-& $0.02^{+0.11}_{-0.11}$ &-& $-0.07^{+0.50}_{-0.50}$ \\
$y_{\rm sub} [\arcsec] - 1.0$ &  & $\mathcal{G}(0.0,0.05)$  &-& $0.03^{+0.12}_{-0.12}$ &-& $-0.05^{+0.38}_{-0.38}$ \\
\hline
 $\theta_E [\arcsec]$ &   &  ${\mathcal{U}(-10,15)}$ & $1.40007^{+0.00028}_{-0.00006}$ & $1.40008^{+0.00030}_{-0.00006}$ & $1.4001^{+0.0006}_{-0.0001}$ & $1.4001^{+0.0006}_{-0.0001}$ \\
$\gamma \in [1.0,3.0]$ &   &  ${\mathcal{G}(2.0,0.2)}$ & $2.20^{+0.01}_{-0.01}$ & $2.15^{+0.13}_{-0.14}$ & $2.24^{+0.54}_{-0.74}$ & $2.06^{+0.21}_{-0.21}$ \\
$x [\arcsec]$ &  &  ${\mathcal{G}(0,0.5)}$ & $0.021^{+0.001}_{-0.001}$ & $0.02^{+0.01}_{-0.01}$ & $0.019^{+0.004}_{-0.004}$ & $0.004^{+0.101}_{-0.101}$ \\
$y [\arcsec]$ &  &  ${\mathcal{G}(0,0.5)}$ &  $0.0392^{+0.0010}_{-0.0010}$ & $0.040^{+0.004}_{-0.004}$ & $0.03^{+0.01}_{-0.01}$ & $0.04^{+0.06}_{-0.06}$ \\
$e_x$ &  &  ${\mathcal{G}(0.0,0.1)}$ & $0.014^{+0.001}_{-0.001}$ & $0.013^{+0.008}_{-0.008}$ & $0.05^{+0.18}_{-0.18}$ & $0.02^{+0.04}_{-0.04}$ \\
$e_y$ & &  ${\mathcal{G}(0.0,0.1)}$ & $-0.047^{+0.002}_{-0.002}$ & $-0.04^{+0.02}_{-0.02}$ & $-0.08^{+0.20}_{-0.20}$ & $-0.05^{+0.06}_{-0.06}$ \\
$\Gamma_1$ &   &  ${\mathcal{U}(-0.2,0.2)}$ & $0.0622^{+0.0008}_{-0.0008}$ & $0.060^{+0.008}_{-0.008}$ & $0.07^{+0.02}_{-0.03}$ & $0.06^{+0.02}_{-0.02}$ \\
$\Gamma_2$ &   &  ${\mathcal{U}(-0.2,0.2)}$ & $-0.0702^{+0.0009}_{-0.0009}$ & $-0.07^{+0.01}_{-0.01}$ & $-0.07^{+0.02}_{-0.02}$ & $-0.12^{+0.29}_{-0.08}$ \\
\hline
$n_{\rm src}$ &   &  ${\mathcal{T}(-1,0.1,10.0)}$ & $0.37^{+0.01}_{-0.01}$ & $0.36^{+0.04}_{-0.03}$ & $0.51^{+0.02}_{-0.02}$ & $0.50^{+0.02}_{-0.02}$ \\
$\zeta_{\rm src}$ &   &  $10^{\mathcal{G}(2.1,1.1)}$ & $21.10^{+0.67}_{-0.65}$ & $20.65^{+1.47}_{-1.37}$ & $25.23^{+0.95}_{-0.92}$ & $21.71^{+28.03}_{-12.23}$ \\
$\sigma_{\rm src}$ &   &  $10^{\mathcal{U}(-5,2)}$  & $0.143^{+0.002}_{-0.002}$ & $0.142^{+0.008}_{-0.008}$ & $0.225^{+0.009}_{-0.009}$ & $0.22^{+0.02}_{-0.02}$ \\
$\alpha_{\rm src}$ &  &  $10^{\mathcal{U}(-1,1)}$ & $1.94^{+0.01}_{-0.01}$ & $1.92^{+0.03}_{-0.03}$ & $2.08^{+0.04}_{-0.04}$ & $2.00^{+0.46}_{-0.39}$ \\
\hline

$\log$ ELBO&&&25206.40&25288.14&25467.91&25502.80\\
$\Delta\log$  ELBO &&&0.00&81.75&261.51&296.40\\
\hline
\end{tabular}

\end{table*}

\section{Correlation Structure of the Gaussian process after the transformation}
\label{sec:GPafterT}

The above described transformation is non-linear in the positions of the grid, but it is possible to show that the resulting field is still a Gaussian process, albeit with a non-stationary kernel, or dense covariance structure.

 To show this, again we consider a function $f(\vec x)$ evaluated at locations $f(\vec x_i)$, such that we can view this function as a vector $\vec f$. We assume that we have some invertible and bijective function $\vec g(\vec x) = \vec y $, such that $\vec g^{-1}(\vec y) = \vec x$. We can define back and forth transformation between two spaces that are stretched and squeezed in some places relative to each other:

\begin{equation}
    h(\vec x) = \int d \vec y ~ f(\vec y) \delta^D \left[  \vec y - \vec g(\vec x) \right]= f(\vec g(x))
\end{equation}
and
\begin{equation}
    f(\vec x) = \int d \vec y ~ h(\vec y) \delta^D \left[  \vec y - \vec g^{-1}(\vec x) \right]= h( \vec g^{-1}(\vec x))= f( \vec g( \vec g^{-1}(\vec x)))\,.
\end{equation}

In the vector convention, we can write the above as a Linear Transformation. We define the vector $\vec f$ by sampling $f$ at the grid points, such that $f_i = f(\vec y_i)$. Absorbing the quadrature weights into the response matrix, $D_{ij} \leftarrow \Delta \vec y_i D_{ij}$, we can write:

\begin{equation}
h_j = \sum_i D_{ij} f_i \,.
\end{equation}
In matrix notation this becomes $\vec h = D^T \vec f$ with $ \vec f = D^{-T}\vec h $ and $ D^{-T} \cdot D^T = (D^T)^{-1} \cdot D^T = \mathbb{I}.$

This approach extends to the limit of infinite pixels. Since this transformation is effectively linear and $\vec f$ follows Gaussian statistics, it follows that $\vec h$ must also follow Gaussian statistics:
\begin{equation}
    \mathcal{P}(\vec f) = \mathcal{G} \left(\vec f, \Sigma\right) ~~~ \rightarrow ~~~ \mathcal{P}(\vec h) =  \mathcal{G}\left (\vec h, {D}^T \Sigma {D}\right ) 
\end{equation}
The main difference between the two random fields is their correlation structure. In fact, using this approach it is possible to obtain more general correlation structures than one that just depends on the distance of points (at least after transformation). We also know that the correlation structure will change such that:
\begin{equation}
\Sigma_{ij} =  \Sigma(\vec x_i, \vec x_j) \rightarrow \Sigma(\vec g(\vec x_i), \vec g(\vec x_j)) = \big[D^T \Sigma D\big]_{ij} \,.
\end{equation}


\section{Fitting the Transformation}
\label{sec:fitting}

In this Section, we provide a brief description for how the transformation is defined.  Our goal is to find a function that takes an arbitrary distribution of points and maps them to a uniform set of points.  If we think of the target distribution as a probability density function (PDF), this transformation function is given by the cumulative distribution function (CDF).

\begin{figure*}
    \centering
    \includegraphics[width=0.7\hsize]{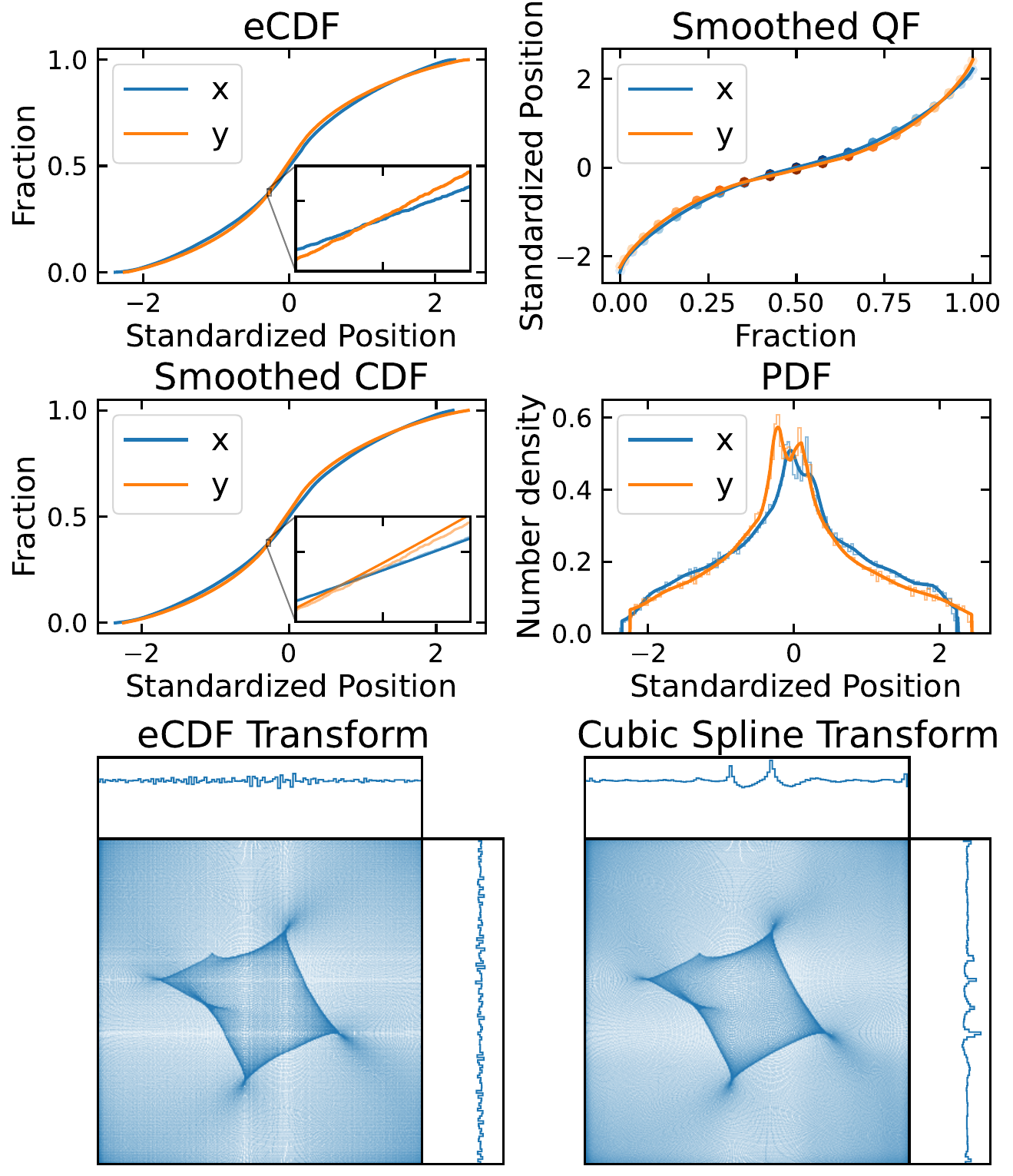} 
    
    \caption{The transformation fitting process.  Upper left: The eCDF for the x (blue) and y (orange) coordinates of mock 1.  Zooming in on this plot (inset) we can see that it is not smooth.  Upper right: The smoothed QF evaluated on a set of Chebyshev nodes (scatter points) color coded by a weight value (darker values have higher weight) derived from the slope of the eCDF at the nodes.  These weighted nodes are used to fit an order 21 polynomial (solid lines).  Middle left: The inverse of this polynomial is approximated with a cubic spline to produce a smoothed CDF.  The inset shows a zoomed in region with the original eCDF (lighter colors) alongside the smoothed CDF (darker colors).  Middle right: The auto derivative of the cubic spline (lines) over-plotted with the projected histogram of the number of rays along each direction (histogram).  The cubic spline transformation produces a smooth approximation to the marginal histograms. Lower left: The positions and marginals of the super-sampled rays after applying a transformation only based on the (non-smoothed) eCDF.  There are clear banding artifacts visible.  Lower right: The positions and marginals of the super-sampled rays after applying a transformation based on the smoothed CDF.  The smoothing process removes the banding artifacts.}
    \label{fig:smooth_fittingA}
    \label{lastpage}
\end{figure*}

To first order, the transformation can be defined by constructing an empirical CDF (eCDF) and using linear interpolation between the observed positions to map any input value on the source plane to the range 0 to 1.  Figure \ref{fig:smooth_fittingA} (lower left) shows the result of transforming the traced super-sampled points using the eCDF directly in this way. Although the marginal distributions are perfectly uniform (by construction), at the points of the caustics, there are visible bands of low density both vertically and horizontally.  This method compensates for the high density points of the caustics in the marginals by reducing the density of points along the axes.  To reduce this banding, a smooth approximation to the eCDF is needed.

To smooth the eCDF we do the following:
i) We trace the observed pixels\footnote{We use the observed pixels rather than the super-sampled pixels to save computational cost both in time and memory usage.} inside the arc mask back to the source plane, subtract off the mean in both the x and y coordinates, and divide by the minimum standard deviation of the x and y coordinates.  These standardized positions remove any shifting or scaling that might arise from a mass-sheet-like transformation. ii) Construct the eCDF in both the x and y directions for the observed pixels inside the arc mask (Figure \ref{fig:smooth_fittingA} upper left).  iii) An odd order polynomial is fit to the quartile function (QF, e.g. the {\textit{inverse}} of the eCDF).  To avoid overfitting and ringing artifacts in this fit, we use a set of Chebyshev nodes equal to the order of the fit.  Each node is assigned a weight that is equal to the slope of the eCDF at that node\footnote{If these were probability distributions, this would be the same as weighting by the value of the PDF at each node.}.  Figure \ref{fig:smooth_fittingA} (upper right) shows these nodes and the resulting fit of a 21st order polynomial. iv) While this polynomial gives the QF, we need the CDF, so this function needs to be inverted.  Numerical inversion of this polynomial could be used, but these tend to be slow or have memory issues for auto differentiation\footnote{Numerical inversion methods are typically implicit functions that refine a guess until some tolerance is reached (e.g. a while loop).  In JAX the reverse-mode auto-derivative of an implicit function would require a program with unbounded memory use.  While it is possible to write a custom auto-derivative rule to avoid these memory issues, the resulting code is slower than if an explicit function for the inverse can be defined.}, so instead we use a cubic spline to approximate the inverse using the Chebyshev nodes from before as the knots for the spline\footnote{We explored fitting a smoothing spline directly to the eCDF but found it to be computationally more expensive than the outlined method}.  The functional form of this spline is computationally quick to calculate, since a cubic spline is fully defined from the values of the function and its derivative at each knot.

Figure \ref{fig:smooth_fittingA} (lower right) shows the resulting distribution and marginals of the traced super-sampled points after using the smoothed eCDF for the transformation to the unit square.  The banding that was visible from using the linearly interpolated eCDF has been eliminated.  Figure \ref{fig:smooth_fittingA} (middle right) shows the projected histogram of rays on the source plane for mock 1 (histogram) along with the auto-differentiated spline smoothed transformation function (lines).  This clearly shows that our transformation creates a smooth estimate of our target density.

\end{document}